\documentclass[aps]{revtex4}
\usepackage{amsfonts}
\usepackage{amsmath}
\usepackage{amssymb,epsf}
\usepackage{color}
\usepackage{graphicx}
\usepackage{epstopdf}
\usepackage{float}
\usepackage{caption}
\usepackage{subfig}

\begin{document}

\title{Effect of massive graviton on dark energy star structure}
\author{A. Bagheri Tudeshki$^{1}$\footnote{%
email address: a.bagheri@hafez.shirazu.ac.ir}, G. H. Bordbar$^{1}$\footnote{%
email address: ghbordbar@shirazu.ac.ir}, and B. Eslam Panah$^{2,3,4}$\footnote{%
email address: eslampanah@umz.ac.ir}}
\affiliation{$^{1}$ Department of Physics and Biruni Observatory, Shiraz University,
Shiraz 71454, Iran \\
$^{2}$ Department of Theoretical Physics, Faculty of Science, University of
Mazandaran, P. O. Box 47415-416, Babolsar, Iran\\
$^{3}$ ICRANet-Mazandaran, University of Mazandaran, P. O. Box 47415-416,
Babolsar, Iran\\
$^{4}$ ICRANet, Piazza della Repubblica 10, I-65122 Pescara, Italy}

\begin{abstract}
The presence of massive gravitons in the field of massive gravity is
considered an important factor in investigating the structure of compact
objects. Hence, we are encouraged to study the dark energy star structure in
the Vegh's massive gravity. We consider that the equation of state governing
the inner spacetime of the star is the extended Chaplygin gas, and then
using this equation of state, we numerically solve the
Tolman-Oppenheimer-Volkoff (TOV) equation in massive gravity. In the
following, assuming different values of free parameters defined in massive
gravity, we calculate the properties of dark energy stars such as radial
pressure, transverse pressure, anisotropy parameter, and other
characteristics. Then, after obtaining the maximum mass and its
corresponding radius, we compute redshift and compactness. The obtained
results show that for this model of dark energy star, the maximum mass and
its corresponding radius depend on the massive gravity's free parameters and
anisotropy parameter. These results are consistent with the observational
data and cover the lower mass gap. We also demonstrate that all energy
conditions are satisfied for this model, and in the presence of anisotropy,
the dark energy star is potentially unstable.
\end{abstract}

\maketitle

\section{Introduction}

\label{sec1}

The existence of singularities in physics cannot be denied. Since a
reliable physical theory must be free from these singularities, the
existence of dark energy (DE) in various models (which include the
cosmological constant $\Lambda $, quintessence \cite{Caldwell1998}, phantom
energy \cite{Caldwell2002}, Chaphygin gas \cite{Kamenshchik2001}, etc.,
which can provide an explanation for the accelerated expansion of the 
Universe) may be an option to resolve this problem in the case of
compact objects. Several attempts were made to provide the models for
compact objects and the existence of dark energy in them, including the
introduction of false vacuum bubbles \cite{Coleman1980}, non-singular black
holes \cite{Dymnikova1992}, gravastars \cite{MazurM2004}, and also dark
energy stars (DES) \cite{ChaplinearXiv}. The final stage of gravitational
collapse of a compact object whose mass is greater than the mass of a
neutron star, according to Chapline's proposal \cite{ChaplinearXiv}, turns
into a DES in which there is no singularity in its center, and the surface
of the star is introduced as a critical surface. In fact, for a
compact object undergoing gravitational collapse, high-frequency quantum
fluctuations exist near where the event horizon is defined. As a result of
these very high-frequency fluctuations, it is expected that all nucleons in
any kind of matter undergoing gravitational collapse will collectively decay
into a mixture of leptons, photons, and vacuum energy droplets. This vacuum
energy displays the difference in the ground state energy between a gas of
quarks and a gas of leptons \cite{Chapline11,Barbieri1,Chapline2014}.
However, these droplets of vacuum energy will be unstable because their mass
is too small to satisfy the de Sitter condition a region of space with
vacuum energy will be stable only if its radius is equal to its
gravitational radius. Consequently, such droplets would immediately expand
and coalesce, forming a uniform vacuum energy. Finally, after the escape of
other decay products, only a compact object remains, whose inner region may
be filled by vacuum energy \cite{Chapline11,Barbieri1,Chapline2014}.

After the introduction of DES, due to having very large vacuum energy in the
center and the occurrence of phase transition on the surface, the presence
of anisotropic pressure in it was investigated in several studies. Also, the
non-establishment of the hydrostatic equilibrium regarding the isotropic
pressure in gravastars \cite{Cattoen2005}, became another reason for the
idea of anisotropy to remain strong in DES. A spherically symmetric model
with a hypersurface was proposed by Lobo \cite{Lobo2006}, in which the
negative radial pressure, positive anisotropy parameter, and negative
gravity profile are the features of DES. For the first time, he investigated
the stability of this star. Ghezzi \cite{Ghezzii2011} introduced a
configuration of the inner spacetime of DES, which contained dark energy as
well as a neutron gas. Unlike Lobo's work \cite{Lobo2006} where the radial
and transverse pressures were equal only at the center of the star, Ghezzi
equated these two pressures inside the star and considered them to be
different only at the surface. He showed that the maximum mass of DES
depends on the coupling parameter. Also, a non-singular stable model of DES
was presented in which the radial pressure was proportional to the matter
density \cite{Rahaman2012}. On the other hand, a combination of baryonic
matter and the phantom scalar field was a model that contained an exact
solution for DES \cite{Yazadjiev2011}. In this way, various models of DES
were presented over time, most of them referring to the anisotropy parameter
and stability of the star. In addition, other definable properties were also
studied for it \cite{BharR2015,Bharetal2018,Banerjee2020,Bhar2021}. Also, it
was shown in Ref. \cite{Sakti2021} that the intervention of the phantom
field can cause the instability of the star. The time dependence of
Einstein's equations and the existence of a negative final pressure for a
DES created a configuration to have no central singularity \cite%
{BeltracchiG2019}. The radial oscillations of DES governed by the extended
equation of state (EoS) of Chaplygin gas were measured by Panotopoulos et
al. \cite{Panotopoulos2020}. Assuming the presence of slow rotations in DES,
measurements for an isotropic sample showed that the moment of inertia of a
rotating star is less than a non-rotating star \cite{Panotopoulos2021}.
Solving the Einstein-Maxwell field equations in a model of Finch-Skea
spacetime \cite{Finch} was done by Malaver \cite{Malaver2022}. He stated
that the physical parameters of a DES behave well in the presence of
electric charge. Recently, for two cases of isotropic and anisotropic DESs,
Pretel \cite{Pretel2023} investigated other physical quantities such as the
radial pulsations and tidal deformation.

One approach in investigating the behavior of compact objects such
as black holes and massive stars is to consider new gravitational fields,
which can be followed in modified gravity. The black hole solution and their
properties including thermodynamic properties, cosmic singularities, and
entropy limit were studied in paradigmatic F(R) models \cite{Dombriz}.
Various studies were carried out, particularly on neutron stars in modified
gravity. Yazadjiev et al. showed that the R-squared gravity parameter is
effective in the neutron star mass-radius relations, however, it is
comparable with other results obtained for the different equations of state 
\cite{Yazadjiev11}. By assuming slowly and rapidly rotating neutron stars in
the R-squared gravity and $R^{2}$ gravity of modified gravity, it was argued
that rotation increases the deviations from general relativity and the
maximum mass and moment of inertia can reach higher values \cite%
{Staykov,Yazadjiev12}. It was shown that in $R^{2}$ gravity in both the
slowly and rapidly rotating modes, the $I-Q$ relation can be distinguished
from the results of general relativity (GR) compared to other modified
gravity models, which can be used as an observational constraint in F(R)
studies \cite{Doneva}. In the context of $R^{2}$ gravity, it was also shown
that the Tidal Love Numbers for non-rotating neutron stars can be several
times larger compared to GR \cite{Yazadjiev13}. The interesting point is
that in $F(R)=R+\alpha R^{2}$ gravity, the maximum mass and compactness have
smaller values compared to GR, which is similar to a repulsive field \cite%
{Feola}. For other $F(R)$ gravity models such as $%
F(R)=R+\alpha R^{2}(1+\gamma R)$, it was indicated that the mass-radius
relation strongly depends on the value of the curvature corrections, the
sign of the correction parameters and the chosen equation of state \cite%
{Capozziello}. In this regard, it was shown in a tensor scalar theory that
for a rotating neutron star, the maximum mass can be greater than the
maximum mass specified in GR \cite{Doneva1}. Another theory of modified
gravity is of interests to models based on inflation. In response to the
question that inflationary fields in various models such as the inflationary
Higgs theory and $R^{p}$ attractor theories can be effective in the
phenomenological description of the neutron star, Oikonomou showed that for
the different equation of state models, the maximum mass obtained in these
theories can increase compared to GR \cite{Oikonomou0,Oikonomou,Oikonomou1}.
To have a general and comprehensive view of other relativistic and
non-relativistic stars structure in different theories of modified gravity,
it is very useful to study reference \cite{Olmo}.

Modified gravity can be important in the field of DES structure in two ways.
First, in line with GR, it can improve the results of modified gravity
compared to GR without changing the concept of dark energy. For example,
Malaver et al. \cite{Malaver2021} showed that the physical properties of a
DES are maintained in Einstein-Gauss-Bonnet gravity. Also, by assuming the
dependence of metric potentials on energy, Tudeshki et al. \cite%
{Tudeshki2022} demonstrated in addition to satisfying the characteristics of
DES, the stability of a star near the surface depends on energy, and the
results obtained in gravity's rainbow are improved compared to GR gravity.
Secondly, the origin of the late acceleration of the \textbf{Universe} is
not yet precisely known. In general, the concept of dark energy and the
cosmological constant try to solve the problem based on the demand of GR.

However, persistent problems in cosmology have led to views of modified
gravitation as an alternative to dark energy. One such theory is called
massive gravity, which was first formulated by Pauli and Fierz \cite%
{PauliFierz}. The introduction of a spin$-2$ mass field in this theory,
especially the assignment of mass to the graviton, became a turning point to
justify the recent accelerated expansion at large distances without needing
the dark energy. However, this linear theory had a fundamental flaw. In the
massless limit of graviton, it does not satisfy GR. A non-linear method was
an idea that Vainshtein \cite{Vainshtein} offered to remove this problem.
But Boulware and Deser (BD) \cite{BD} showed that the existence of the ghost
in this non-linear theory becomes another problematic factor. In order to
solve this problem, there were many studies that have led to the formation
of sub-branches in massive gravity. For example, the new massive gravity
(NMG) in three dimensions \cite{Bergshoeff2009}. In 2009 and 2010 in two
studies \cite{deRham2010,deRham2011}, it was shown that by using the
graviton's mass and polynomial interaction terms involved in spacetime
action, BD-ghost can be prevented. This is possible by introducing a
reference metric. This method is known as dRGT massive gravity. Providing an
exact static solution of the black hole in this gravity \cite{Ghosh2016}
allowed the study of compact objects in the presence of dRGT massive gravity
to be strengthened. In particular, the presence of two additional terms in
the massive metric coefficients compared to GR gravity can be a
justification for dark matter and dark energy in massive gravity. It can be
seen from this point of view that the massive graviton can play the role of
the cosmological constant in cosmic distances \cite%
{Gumrukcuoglu,Comelli,Langlois,Kobayashi}. Also, the multi-gravity theory
has tried to solve the BD-ghost problem in multi-dimensional spacetime for
interacting spin-$2$ fields by applying the vielbein formulation \cite%
{Hinterbichler}. In another interesting study, Hassan and Rosen showed that
by using a dynamic reference metric instead of a static reference metric,
the BD-ghost can be neglected in a non-linear bimetric theory for a massless
spin-$2$ field \cite{Hassan and Rosen}. A branch of dRGT massive gravity
theory was proposed in 2013 by Vegh \cite{VegharXiv}. By applying
holographic principles and a singular reference metric, he was able to
establish a ghost-free theory. After the introduction of Vegh's model, many
works were done in this field. Hendi et al., \cite%
{Hendi2016a,Hendi2016b,Hendi2016c,Hendi2016d,Hendi2016f,Hendi2016g} studied
the stability and thermodynamics of black holes by investigating the effect
of massive parameters. Also, in Ref. \cite{Hendi2017}, they investigated the
physical quantities of the neutron star, including the maximum mass, using
the AV18 potential method. Eslam Panah and Liu \cite{PanahLiu} demonstrated
that the maximum mass of a white dwarf in the presence of this gravity is
greater than the Chandrasekhar limit. It was indicated that the end state of
Hawking evaporation led to a black hole remnant in Vegh's massive gravity,
which could help to ameliorate the information paradox \cite{PanahHY}. Also,
there was a correspondence between black hole solutions of conformal and
Vegh's massive theories of gravity \cite{PanahH}. The remarkable point in
Vegh's massive gravity is that the metric coefficients of spacetime do not
have a term with the square of the distance $r^{2}$. In the sense that
cosmological constant $\Lambda $ can be defined for it separately. In
addition, in the definition of a DES \cite{ChaplinearXiv}, we are dealing
with dark energy that has a much greater density than the cosmological
constant. Therefore, in the present study, we investigate the physical
characteristics of this compact object by using Vegh's massive gravity and
introducing dark energy in the form of fluid inside the inner spacetime.

In the last two decades, various observational and theoretical methods were
introduced to constrain the range of graviton's mass. Among others, we can
mention the observation of gravitational waves emitted by the merger of a
binary system, which was carried out by the LIGO-Virgo collaboration. In
this measurement, the range of graviton's mass in the GW170104 event was
determined to be $m_{g}<7.7\times 10^{-23}eV$ \cite{Scientific2018}. To
learn about other methods, one can see Refs. \cite{Goldhaber,Berti}. On the
other side of the theory, the idea is raised as to why should the mass of
the graviton be constant. Attributing variable mass dependent on the scalar
field is the work that was done in response to this question by Huang et
al., \cite{Huang}, which is known as mass-varying massive gravity (MVMG).
Following that, a number of additional studies have been conducted,
including investigating the effect of the massive graviton in strong and
weak gravitational environments. The authors \cite{Zhang2018} suggested that
the mass of graviton near a black hole could be $10^{23}eV$ orders of
magnitude greater than that in the presence of weak gravitational fields. 
In this study, it was shown how, regardless of the presence of
exotic matter or quantum effects, the mass of the graviton, which is
approximately of Hubble scale $10^{-33}eV$ in weak gravitational
environments, can increase to approximately, $10^{-10}eV$ near the event
horizon of a black hole under environmental effects. Especially, it appears
near the horizon and alters the GWs resulting from black hole mergers by
producing "echoes" at the end time \cite{Cardoso,Cardoso1}. Also, in
another study \cite{Sun}, these effects on neutron stars and white dwarfs
were checked, which showed that the results agreed with the Ref. \cite%
{Zhang2018} reported an increase in the mass of graviton near the star.

In the present work, in order to compare the behavior of DES in the presence
of massive gravitons with the observational results obtained so far, we use
the observational measurement of the GW190814 event \cite{Abbott2020} and
the mass-radius relation for NS pulsars J1614-2230 \cite{Demorest2010},
J0348+0432 \cite{Antoniadis2013}, J0740+6620 \cite{Cromartie2020} and
J2215+5135 \cite{Linares2018}. We also use the observational results of a
binary system including the giant 2MASS J05215658+4359220 and a massive
unseen companion \cite{Thompson2019}.

In this paper, we follow the scheme of this study as follows: in Sec. \ref%
{sec2}, we take a look at the background and relations of Vegh's massive
gravity. Then, considering a spherically symmetric spacetime in Sec. \ref%
{sec3}, we obtain the equations of motion and by using them, we introduce
the TOV equation for the anisotropic distribution in Vegh's massive gravity.
In this section and in the following, we present EoS and a model for the
existence of anisotropy. In Sec. \ref{sec4}, by numerically solving the
modified TOV equation, we obtain and analyze the properties of DES, such as
maximum mass and its corresponding radius, radial pressure and transverse
pressure, anisotropy parameter, surface redshift, etc. Finally, in Sec. \ref%
{sec5}, we present a summary of the results obtained from this study.

\section{Massive Gravity}

\label{sec2} In ghost free massive gravity, the action is given by \cite%
{deRham2011} 
\begin{equation}
I=\frac{c^{4}}{16\pi G}\int d^{4}x\sqrt{-g}[\mathcal{R}+m_{g}^{2}{%
\sum\limits_{i}^{4}c_{i}\mathit{u_{i}(g,f)}}]+I_{matter},
\end{equation}%
where $c$ and $G$ are the speed of light and the gravitational constant,
respectively. Also, $\mathcal{R}$ is the Ricci scalar and $m_{g}$ refers to
the graviton mass. In the second term on the right side of the action
(potential term), $c_{i}$ are constant coefficients that play the role of
free parameters of action. Also $u_{i}$ are introduced as symmetric
polynomials of the eigenvalues of matrix $K_{~~b}^{a}=\sqrt{g^{ac}f_{cb}}$.
Here $g_{ab}$ is dynamical metric tensor and $f_{ab}$ is called the
reference metric. At the end of the above action, $I_{matter}$ implies the
action of matter. $u_{i}$ are expressed in the following form 
\begin{equation}
{\mathit{u}}_{i}=\sum_{y=1}^{i}\left( -1\right) ^{y+1}\frac{\left(
i-1\right) !}{\left( i-y\right) !}{\mathit{u}}_{i-y}\left[ K^{y}\right] ,
\label{U}
\end{equation}%
where ${\mathit{u}}_{i-y}=1$, when $i=y$. Note; in the above relations, the
bracket marks indicate the traces in the form; $[K]=K_{a}^{a}$ and $%
[K^{n}]=(K^{n})_{a}^{a}$. Here we intend to specify the equations of motion
in massive gravity, thus by varying the action with respect to metric tensor 
$g_{ab}$, and after doing some calculations, the equations are obtained as
follows 
\begin{equation}
G_{ab}+m_{g}^{2}\chi _{ab}=\frac{8\pi G}{c^{4}}T_{ab},  \label{field eq}
\end{equation}%
where $G_{ab}$ is Einstein tensor, $\chi _{ab}$ is called massive tensor and 
$T_{ab}$ is stress-energy tensor. Note; in the following, we use geometrized
units $(c=G=1)$ for calculations. The massive tensor $\chi _{ab}$ is
extracted to form 
\begin{equation}
\chi _{ab}=-\sum_{i=1}^{d-2}\frac{c_{i}}{2}\left[ {\mathit{u}}%
_{i}g_{ab}+\sum_{y=1}^{i}\left( -1\right) ^{y}\frac{i!}{\left( i-y\right) !}{%
\mathit{u}}_{i-y}\left[ K_{ab}^{y}\right] \right] ,  \label{chi}
\end{equation}%
where $d$ is related to the dimensions of spacetime. We work on
4-dimensional spacetime and so $d=4$.

\section{Equations of motion and hydrostatic equilibrium equations}

\label{sec3} We consider a spherically symmetric spacetime with metric
signature $(-,+,+,+)$ 
\begin{equation}
ds^{2}=-e^{2\nu }dt^{2}+e^{2\lambda }dr^{2}+r^{2}(d\theta ^{2}+\sin
^{2}(\theta )d\phi ^{2}),  \label{metric}
\end{equation}%
where $e^{2\nu }$ and $e^{2\lambda }$ are the metric potentials. The spatial
reference metric or spatial fiducial metric is a suitable option for the
black hole solutions \cite{VegharXiv,Cai,Hendi2017,PanahLiu}. Hence we
suppose 
\begin{equation}
f_{ab}=diag\left( 0,0,C^{2},C^{2}\sin ^{2}\theta \right) ,
\label{reference metric}
\end{equation}%
where $C$ is a positive constant. Using the line element Eq. (\ref{metric})
and reference metric Eq. (\ref{reference metric}), we can determine the
tensor $K_{~~b}^{a}$ 
\begin{equation}
K_{~~b}^{a}=diag\left( 0,0,\dfrac{C}{r},\dfrac{C}{r}\right) ,
\label{k tensor}
\end{equation}%
and 
\begin{eqnarray}
(K^{2})_{~~b}^{a} &=&\left[ 
\begin{array}{cccc}
0 & 0 & 0 & 0 \\ 
0 & 0 & 0 & 0 \\ 
0 & 0 & \frac{{C}^{2}}{{r}^{2}} & 0 \\ 
0 & 0 & 0 & \frac{{C}^{2}}{{r}^{2}}%
\end{array}%
\right] ,  \notag \\
&&  \notag \\
(K^{3})_{~~b}^{a} &=&\left[ 
\begin{array}{cccc}
0 & 0 & 0 & 0 \\ 
0 & 0 & 0 & 0 \\ 
0 & 0 & \frac{{C}^{3}}{{r}^{3}} & 0 \\ 
0 & 0 & 0 & \frac{{C}^{3}}{{r}^{3}}%
\end{array}%
\right] ,  \notag \\
&&  \notag \\
(K^{4})_{~~b}^{a} &=&\left[ 
\begin{array}{cccc}
0 & 0 & 0 & 0 \\ 
0 & 0 & 0 & 0 \\ 
0 & 0 & \frac{{C}^{4}}{{r}^{4}} & 0 \\ 
0 & 0 & 0 & \frac{{C}^{4}}{{r}^{4}}%
\end{array}%
\right] ,  \label{k power}
\end{eqnarray}
and also 
\begin{eqnarray}
\lbrack K] &=&\frac{2C}{r},~~~\&~~~[K^{2}]=\dfrac{2C^{2}}{r^{2}},  \notag \\
&&  \notag \\
\lbrack K^{3}] &=&\dfrac{2C^{3}}{r^{3}},~~~\&~~~[K^{4}]=\dfrac{2C^{4}}{r^{4}}%
.  \label{k trace}
\end{eqnarray}%
Since we are working on 4-dimensional spacetime, the only non-zero terms of $%
u_{i}$ are $u_{1}$ and $u_{2}$\thinspace\ \cite{Hendi2016g}. Therefore,
using Eqs. (\ref{U}), (\ref{k tensor}) and (\ref{k trace}), $u_{i}$ are
obtained 
\begin{eqnarray}
u_{1} &=&\dfrac{2C}{r},~~~\&~~~u_{2}=\dfrac{2C^{2}}{r^{2}},  \notag \\
u_{i} &=&0,\text{ \ \ when \ \ }i>2.  \label{ui}
\end{eqnarray}%
By putting Eq. (\ref{ui}) in Eq. (\ref{chi}), the elements of the massive
tensor are determined 
\begin{eqnarray}
\chi _{~~1}^{1} &=&-\dfrac{C(Cc_{2}+c_{1}r)}{r^{2}},~~~\&~~\chi _{~~3}^{3}=-%
\dfrac{c_{1}C}{2r},  \notag \\
&&  \notag \\
~\chi _{~~2}^{2} &=&-\dfrac{C(Cc_{2}+c_{1}r)}{r^{2}},~~~\&~~~\chi
_{~~4}^{4}=-\dfrac{c_{1}C}{2r}.  \label{chi1}
\end{eqnarray}

We consider that the interior of the star is filled with an anisotropic
fluid. The stress-energy tensor for an anisotropic distribution $T_{ab}$, is
applied according to the following definition \cite{Bayin1986} 
\begin{equation}
T_{ab}=\left[ \rho \left( r\right) +p_{t}\left( r\right) \right]
v_{a}v_{b}+p_{t}\left( r\right) g_{ab}+\left[ p_{r}\left( r\right)
-p_{t}\left( r\right) \right] x_{a}x_{b},  \label{T4}
\end{equation}%
where $\rho $, $p_{r}$, and $p_{t}$ are the energy density, the radial
pressure, and the transverse pressure, respectively. Also $v_{a}$ refers to
the four-velocity vector with $v_{a}v^{a}=-1$, and $x_{a}$ represents the
unit spacelike vector with $x_{a}x^{a}=1$. According to the metrics of the
line element (\ref{metric}), the diagonal elements of the stress-energy
tensor are obtained: 
\begin{equation}
T_{1}^{1}=-\rho
(r),~~~\&~~~T_{2}^{2}=p_{r}(r),~~~\&~~~T_{3}^{3}=T_{4}^{4}=p_{t}(r).
\end{equation}
By substituting the above equations and the set of Eqs. (\ref{chi1}) in Eq. (%
\ref{field eq}), the equations of motion in massive gravity are obtained 
\begin{eqnarray}
\dfrac{2r\lambda ^{\prime -2\lambda (r)}-e^{-2\lambda
(r)}+1+m_{g}^{2}C(c_{2}C+c_{1}r)}{r^{2}} &=&8\pi \rho (r),  \label{GR1} \\
&&  \notag \\
\dfrac{2r\nu ^{\prime -2\lambda (r)}+e^{-2\lambda
(r)}-1-m_{g}^{2}C(c_{2}C+c_{1}r)}{r^{2}} &=&8\pi p_{r}(r),  \label{GR2} \\
&&  \notag \\
e^{-2\lambda (r)}\left[ \nu ^{\prime \prime }(r)-\lambda ^{\prime }(r)\nu
^{\prime }(r)+\nu ^{\prime 2}(r)-\frac{\lambda ^{\prime }(r)}{r}+\frac{\nu
^{\prime }(r)}{r}\right] -\frac{c_{1}m_{g}^{2}C}{2r} &=&8\pi p_{t}(r),
\label{GR3}
\end{eqnarray}%
where the prime and double prime are representing the first and second
derivatives with respect to $r$, respectively. By moving the
graviton mass terms (the last term on the right side of the Eqs. (\ref{GR1})
and (\ref{GR2})) to the other side of equality, one can see that the massive
gravitons are similar to a fluid which has pressure and density. Another
interesting point is that according to Eqs. (\ref{GR2}) and (\ref{GR3}), we
realize that the pressure caused by gravitons is anisotropic. This
anisotropic fluid can pattern the behavior of the dark matter halo on large
scale \cite{Panpanich2018}.

The first equality of the equations of motion (\ref{GR1}) leads us to the
following relation 
\begin{equation}
e^{-2\lambda (r)}=1-\dfrac{2m(r)}{r}+m_{g}^{2}C\left( \dfrac{c_{1}r}{2}%
+c_{2}C\right) ,  \label{lambda}
\end{equation}%
and 
\begin{equation}
m(r)=\int_{0}^{r}4\pi r^{\prime 2}\rho \left( r^{\prime }\right) dr^{\prime
},  \label{mass}
\end{equation}%
where $m\left( r\right) $ shows the mass-energy function. Using the second
equality (\ref{GR2}), we obtain the gravity profile relation, 
\begin{equation}
g(r)=\frac{d\nu (r)}{dr}=\dfrac{Cc_{1}m_{g}^{2}r^{2}+16\pi p_{r}r^{3}+4m(r)}{%
2r\left( 2C^{2}c_{2}m_{g}^{2}r+Cc_{1}m_{g}^{2}r^{2}-4m(r)+2r\right) }.
\label{nu}
\end{equation}

The conservation condition must be satisfied. So, we have $\triangledown
^{a}T_{ab}=0$. Now, by putting Eq. (\ref{nu}) in the conservation relation,
and solving it in terms of the radial pressure gradient, the hydrostatic
equilibrium equation in Vegh's massive gravity for an anisotropic
distribution is obtained in the following form, 
\begin{equation}
\frac{dp_{r}\left( r\right) }{dr}=\dfrac{\left[ \left( 4\pi
r^{3}p_{r}(r)+m(r)\right) +Cc_{1}m_{g}^{2}r^{2}/4\right] \left( \rho
(r)+p_{r}(r)\right) }{-r\left(
Cc_{1}m_{g}^{2}r^{2}/2-2m(r)+r(m_{g}^{2}c_{2}C^{2}+1)\right) }+\dfrac{%
2\left( p_{t}(r)-p_{r}(r)\right) }{r}.  \label{ModTOVI}
\end{equation}

In the next step, we must introduce the EoS that can show the behavior of
dark energy, and a model to describe the fluid anisotropy in order to solve
the modified TOV equation. A suitable choice is generalized Chaplygin gas
EoS $p=-\dfrac{B}{\rho ^{\omega }}$, plus a linear term, which is written as
follows \cite{Debnath2004,Pourhassan2013}, 
\begin{equation}
p_{r}=\hat{A}\rho -\frac{\hat{B}}{\rho ^{\omega }},  \label{EOS}
\end{equation}%
where $\hat{A}$ is a positive dimensionless constant, and $\hat{B}$ is a
positive dimension constant in $[L]^{-4}$ units and the parameter $\omega $
is in the range $(0,1]$. The first term is related to a barotropic fluid,
and the second term refers to the generalized form of the Chaplygin gas EoS 
\cite{Bento2002,Gorini2003,Xu2012}. Such an EoS is an alternative to phantom
and quintessence models in dark energy theory. We can consider the EoS (\ref%
{EOS}) with $\omega =1$ in the following form for convenience in
calculations, \cite{Tello2020,Panotopoulos2020,Panotopoulos2021} 
\begin{equation}
p_{r}=A^{2}\rho -\dfrac{B^{2}}{\rho }.  \label{EOS1}
\end{equation}

In this study, we have investigated the possibility of observing
dark energy stars made of a fluid with a negative pressure $\left(-
\dfrac{B^{2}}{\rho }\right) $ and a barotropic component $\left(A^{2}\rho\right) $. The presence of barotropic term makes
it possible to look at the system in such a way that it contains two fluids.
It can be stated that in the presence of matter (regardless of its type and
interactions), dark energy also exists inside the compressed body for some
reasons. As a proposal, this model of dark energy stars may occupy somewhere
between neutron stars and black holes. It is notable that, regardless of how
the EoS of matter is defined, several studies have shown how the presence of
dark energy affects the structure of other stars for example see Ref. \cite%
{Astashenok1}. In order to compare the results obtained from DES models
with the extended Chaplygin EoS in GR \cite%
{Panotopoulos2020,Panotopoulos2021} and massive gravity (our suggestion),
here we consider the constants $A$ and $B$ of the EoS as $A=\sqrt{0.4}$ and $%
B=0.23\times 10^{-3}km^{-2}$. It should be noted that for the causality
condition to be established at the star surface, the radial speed of sound $%
V_{sr}$ must be satisfied in the range $0<V_{sr}^{2}(R)<1$. Consequently, $%
A^{2}<0.5$ \cite{Pretel2023}. In this way, the permissible range for
choosing constant values of $A$ is determined. By identifying the values of $%
A$ and keeping the considerations of density and numerical solution, $B$ can
be selected.

The anisotropy parameter $\Delta $ represents the difference between
transverse pressure and radial pressure, $p_{t}-p_{r}$. The main cause of
anisotropy in compact objects is the presence of densities greater than the
nuclear density, which can be due to the presence of condensation \cite%
{Hartle}, phase transition \cite{Sokolov} or electromagnetic fields \cite%
{Usov}. It is important to determine which models can be proposed to
correctly describe the anisotropy behavior (which is the difference between
radial and transverse pressure inside the star). In this regard, several
models for anisotropy have been proposed in various studies \cite%
{Bowers1974,Cosenza1981,Horvat2010,Doneva2012,Herrera2013,Raposo2019}. The
model that we extract in the present work is a nonlinear anisotropy that was
first proposed by Bowers and Liang \cite{Bowers1974} in general form and is
expressed as follows, 
\begin{equation}
\Delta =p_{t}-p_{r}=\dfrac{\gamma \left( \rho +3p_{r}\right) \left( \rho
+p_{r}\right) r^{n}}{1-\frac{2m}{r}},  \label{delta}
\end{equation}%
where the constant $\gamma $ measures the degree of anisotropy, and the
constant $n$ is greater than $1$. Since in a realistic model of compact
objects $(M/R)_{critical}>0$, the allowed range is $\gamma \leq 2/3$. In
order to eliminate the anisotropy in the center of the star, we must
consider $n=2$. As a result, the final model of anisotropy parameter that we
use in this study is defined as follows, 
\begin{equation}
\Delta =\dfrac{\gamma \left( \rho +3p_{r}\right) (\rho +p_{r})r^{2}}{1-\frac{%
2m}{r}}.  \label{delta1}
\end{equation}%
Therefore, our proposed model in this work for a dark energy star is
introduced by extended Chaplygin gas Eq. (\ref{EOS1}) and the anisotropy
parameter Eq. (\ref{delta1}), which is formulated in the framework of
massive gravity. For $\gamma =0$ and $m_{g}=0$, the results describe an
isotropic DES in GR that depends only on the proposed model for the EoS \cite%
{Panotopoulos2020,Panotopoulos2021}.

\section{Properties of dark energy stars in massive gravity}

\label{sec4}

\subsection{Numerical Solutions}

The star structure is determined by solving three coupled differential
equations, Eqs. (\ref{mass}), (\ref{nu}), (\ref{ModTOVI}), and using the
governing EoS (\ref{EOS1}) and anisotropy parameter (\ref{delta1}). At the
center of the star $r=0$, the initial conditions are $m(r=0)=0$ and $\rho
(r=0)=\rho _{c}$, where $\rho _{c}$ is the central energy density. At the
surface of the star, the radial pressure is negligible. Therefore, the
boundary conditions on the surface $r=R$ are $p_{r}(R)=0$, $m(r=R)=M$ and $%
\nu (R)=\dfrac{1}{2}ln\left( 1-\dfrac{2M}{R}+m_{g}^{2}C\left( \dfrac{c_{1}R}{%
2}+c_{2}C\right) \right) $, where $M$ is the total mass of DES. After
integration from the center $r=0$ to the surface of the star, the radius $R$
and mass $M$ are obtained. Also, due to the dependence of $M$ and $R$ to $%
\rho _{c}$ (or $p_{r,c}$ ), by considering different values of central
energy density (or central radial pressure), we will be able to determine
the maximum mass for DES models. In the numerical solution, it is important
to pay attention to the fact that the anisotropy parameter is zero in the
center ($\Delta (r=0)=0$), and the energy density on the surface is $\rho
_{s}=B/A$. In addition, Eqs. (\ref{mass}) and (\ref{ModTOVI}) contain terms $%
m_{g}^{2}c_{1}$, $m_{g}^{2}c_{2}$, and constant $C$ that show the dependence
of the solution on the mass of graviton and the free parameters. Here we
consider the mass of graviton to be $1.78\times 10^{-65}g$ \cite%
{Ali2016,Hendi2017,PanahLiu}. The results obtained from the numerical
solution are presented for the central density $\varrho _{c}=1.34\times
10^{15}~g/cm^{3}$, and the central pressure $p_{c}=4.33\times
10^{15}~dyn/cm^{2}$ in Tables. \ref{tab1}-\ref{tab3}. 
\begin{table}[h]
\caption{The properties of dark energy star in massive gravity for $%
C=10^{-3} $, $m_{g}^{2}c_{2}=-2\times 10^{-2}$, $\protect\gamma =0.1$, $A=%
\protect\sqrt{0.4}$ and $B=0.23\times 10^{-3}$.}
\label{tab1}
\begin{center}
\begin{tabular}{|c|c|c|c|c|}
\hline\hline
$m_{g}^{2}c_{1}$ ~~ & $M_{max}(M_{_{\odot}})$~~ & $R$ ($km$)~~ & $\sigma$ ~~
& $z_{s}$ \\ \hline\hline
$3\times10^{-1}$ & $2.55$ & $12.36$ & $0.61$ ~ & $0.60$ \\ \hline
$3\times10^{-2}$ & $2.71$ & $12.71$ & $0.63$ ~ & $0.65 $ \\ \hline
$3\times10^{-3}$ & $2.72$ & $12.83$ & $0.63$ ~ & $0.64 $ \\ \hline
$3\times10^{-4}$ & $2.73$ & $12.84$ & $0.63$ ~ & $0.64 $ \\ \hline\hline
$-3\times10^{-1}$ & $2.95$ & $13.31$ & $0.66$ ~ & $0.71$ \\ \hline
$-3\times10^{-2}$ & $2.75$ & $12.87$ & $0.63$ ~ & $0.65$ \\ \hline
$-3\times10^{-3}$ & $2.73$ & $12.84$ & $0.63$ ~ & $0.64$ \\ \hline
$-3\times10^{-7}$ & $2.73$ & $12.84$ & $0.63$ ~ & $0.64$ \\ \hline\hline
\end{tabular}%
\end{center}
\end{table}
\begin{table}[h]
\caption{The properties of dark energy star in massive gravity for $%
m_{g}^{2}c_{1}=-3\times10^{-1}$, $m_{g}^{2}c_{2}=-2\times10^{-2}$, $\protect%
\gamma=0.1$, $A=\protect\sqrt{0.4}$ and $B=0.23\times10^{-3}$.}
\label{tab2}
\begin{center}
\begin{tabular}{|c|c|c|c|c|}
\hline\hline
$C$ ~~ & $M_{max}(M_{_{\odot}})$~~ & $R$ (km)~~ & $\sigma$~~ & $z_{s}$ \\ 
\hline\hline
$10^{-3}$ & $2.95$ & $13.31$ & $0.66$ ~ & $0.71$ \\ \hline
$10^{-4}$ & $2.75$ & $12.87$ & $0.63$ ~ & $0.65$ \\ \hline
$0.5\times10^{-4}$ & $2.74$ & $12.85$ & $0.63$ ~ & $0.64$ \\ \hline
$8\times10^{-5}$ & $2.74$ & $12.86$ & $0.63$ ~ & $0.65$ \\ \hline
$2\times10^{-5}$ & $2.73$ & $12.84$ & $0.63$ ~ & $0.64$ \\ \hline
$2\times10^{-6}$ & $2.73$ & $12.84$ & $0.63$ ~ & $0.64$ \\ \hline\hline
\end{tabular}%
\end{center}
\end{table}
\begin{table}[h]
\caption{The properties of dark energy star in massive gravity for $%
C=10^{-3} $, $m_{g}^{2}c_{1}=-3\times10^{-1}$, $\protect\gamma=0.1$, $A=%
\protect\sqrt{0.4}$ and $B=0.23\times10^{-3}$.}
\label{tab3}
\begin{center}
\begin{tabular}{|c|c|c|c|c|}
\hline\hline
$m_{g}^{2}c_{2}$ ~~ & $M_{max}(M_{_{\odot}})$~~ & $R$ (km)~~ & $\sigma$~~ & $%
z_{s}$ \\ \hline\hline
$-2\times10^{-4}$ & $2.95$ & $13.31$ & $0.66$~ & $0.71 $ \\ \hline
$-3\times10^{-3}$ & $2.95$ & $13.31$ & $0.66$ ~ & $0.71$ \\ \hline
$-4\times10^{-2}$ & $2.95$ & $13.31$ & $0.66$~ & $0.71 $ \\ \hline
$-5\times10^{-1}$ & $2.95$ & $13.31$ & $0.66$~ & $0.71 $ \\ \hline
$-7\times10^{-1}$ & $2.95$ & $13.31$ & $0.66$~ & $0.71 $ \\ \hline
$-9\times10^{-1}$ & $2.95$ & $13.31$ & $0.66$~ & $0.71 $ \\ \hline\hline
\end{tabular}%
\end{center}
\end{table}

\subsection{Pressure, density and anisotropy}

\begin{figure}[tbh]
\centering
\includegraphics[width=0.4\textwidth]{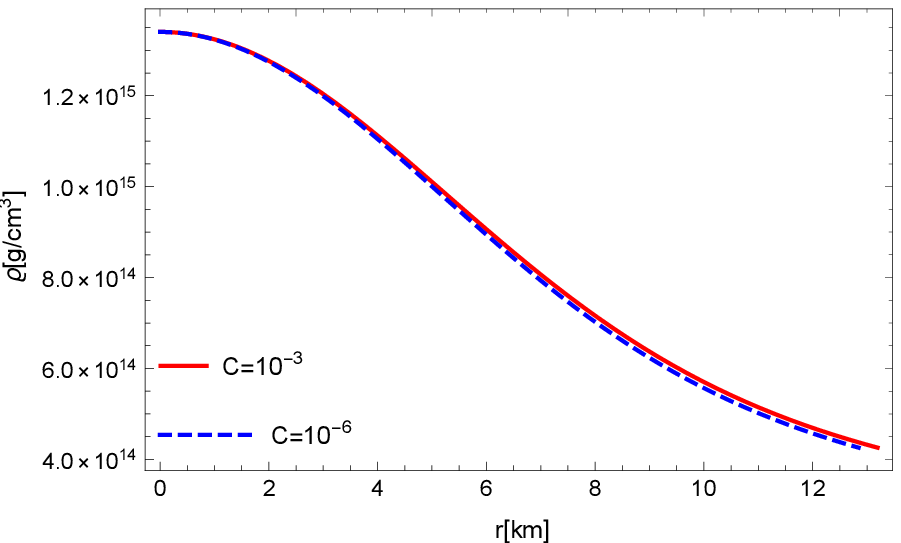} \includegraphics[width=0.4%
\textwidth]{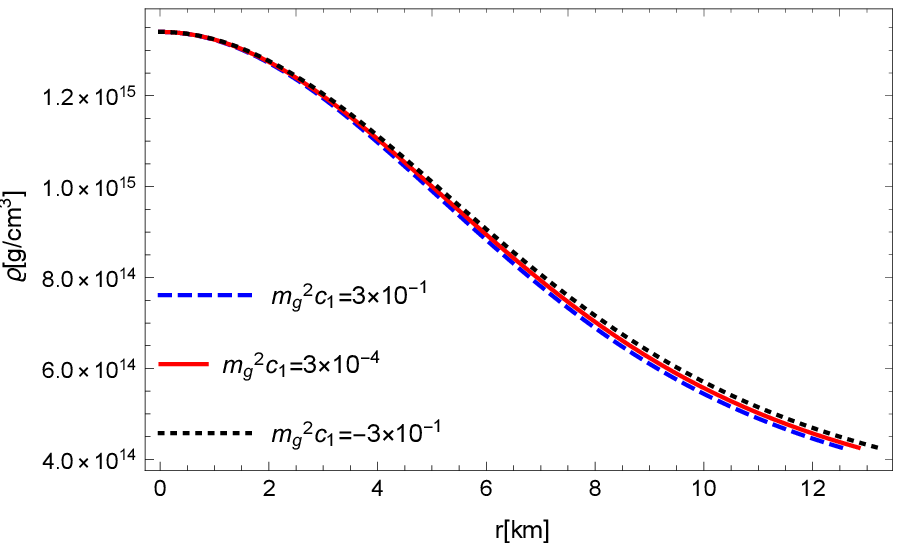} \newline
\caption{ Density $\protect\varrho $ vs radius $r$ for different $C$ (left
panel) and for different $m_{g}^{2}c_{1}$ (right panel) with $%
m_{g}^{2}c_{2}=-2\times 10^{-2}$, $\protect\gamma =0.1$, $A=\protect\sqrt{0.4%
}$ and $B=0.2\times 10^{-3}$. }
\label{Fig1}
\end{figure}

\begin{figure}[tbh]
\centering
\includegraphics[width=0.4\textwidth]{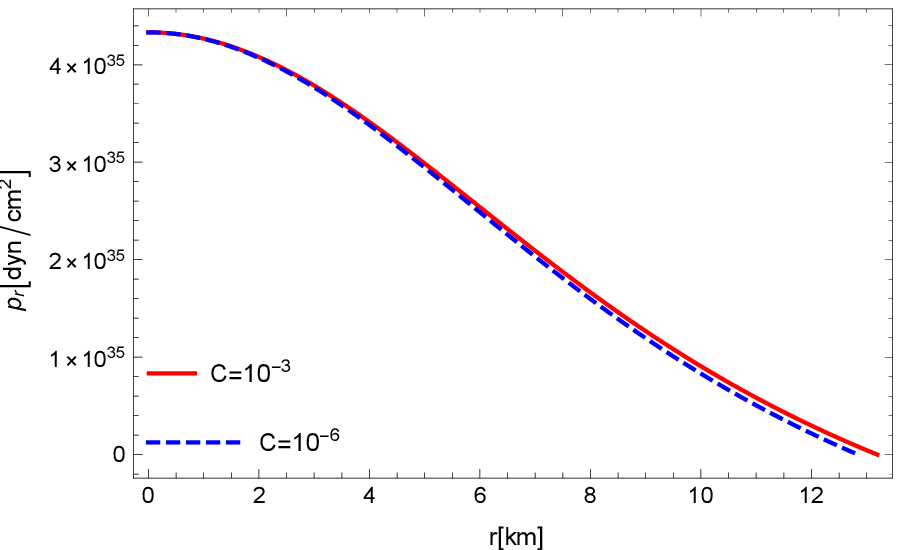} \includegraphics[width=0.4%
\textwidth]{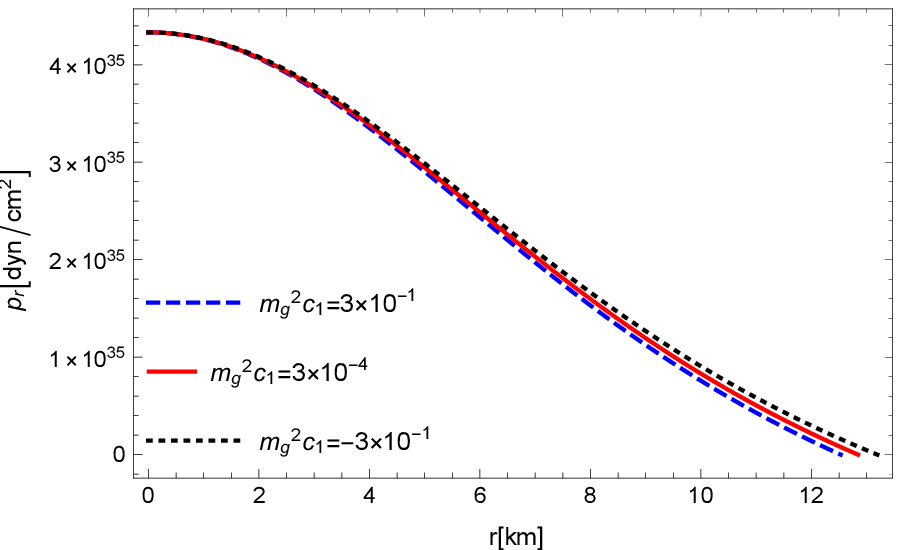} \newline
\caption{ Radial pressure $p_{r}$ vs radius $r$ for different $C$ (left
panel) and for different $m_{g}^{2}c_{1}$ (right panel) with $%
m_{g}^{2}c_{2}=-2\times 10^{-2}$, $\protect\gamma =0.1$, $A=\protect\sqrt{0.4%
}$ and $B=0.2\times 10^{-3}$. }
\label{Fig2}
\end{figure}

\begin{figure}[tbh]
\centering
\includegraphics[width=0.4\textwidth]{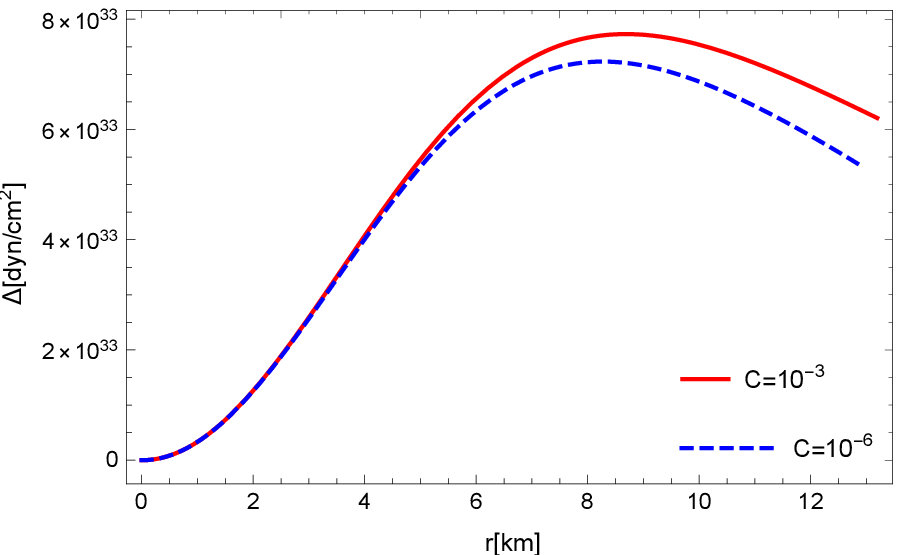} \includegraphics[width=0.4%
\textwidth]{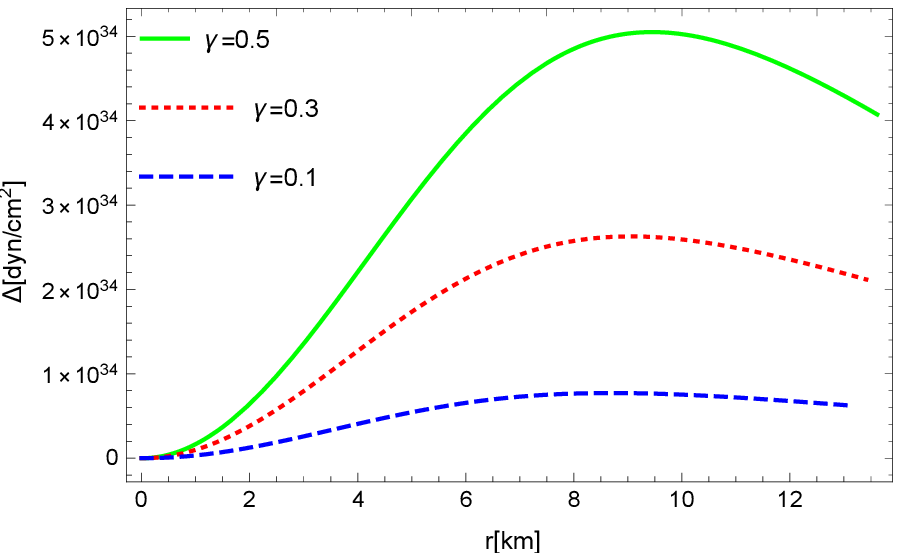} \newline
\caption{Anisotropy parameter $\Delta $ vs radius $r$ for different $C$ and $%
\protect\gamma =0.1$ (left panel) and for $C=10^{-3}$, $m_{g}^{2}c_{1}=-3%
\times 10^{-1}$, $m_{g}^{2}c_{2}=-2\times 10^{-2}$, $A=\protect\sqrt{0.4}$, $%
B=0.2\times 10^{-3}$ and different $\protect\gamma $ (right panel). }
\label{Fig3}
\end{figure}

According to the Eqs. (\ref{mass}), (\ref{ModTOVI}), (\ref{EOS}), (\ref{EOS1}%
) and (\ref{delta1}), and by using a numerical solution, the density, radial
pressure, and anisotropy parameter versus the radius are plotted in the
Figs. \ref{Fig1}-\ref{Fig3}, respectively. Figs. \ref{Fig1} and \ref{Fig2}
show that the density and radial pressure are a decreasing function of the
radius, as is expected. In both plots, it can be seen that the different
values of $C$, and $m_{g}^{2}c_{1}$ affect the behavior of the density and
radial pressure. It should be also noted that different values of $%
m_{g}^{2}c_{2}$ have no effect on changes in these quantities. According to
the left panel of Fig. \ref{Fig3}, it can be seen that the anisotropy
parameter $\Delta $ is also sensitive to the parameter $C$, and increases by
its increasing. Therefore, the massive graviton and the dependence of its
behavior on free parameters can affect the results obtained from density,
pressure, and anisotropy. The factor $\Delta /r$ creates a positive force in
the outward direction of the star. This is an important feature of a DES,
because it ensures the stability of the star. 

It is interesting to examine the change in the behavior of the anisotropy
parameter by changing the value of parameter $\gamma $ versus radius. In the
right panel of Fig. \ref{Fig3}, the $\Delta $ diagram is drawn in terms of
different values of $\gamma $. It can be seen that by increasing $\gamma $,
the amount of anisotropy also increases. This means that the force caused by
it also increases towards the outside of the star, and the star will be more
resistant to gravitational collapse.

%
%

\subsection{Maximum mass, compactness and surface redshift}

Studying the structure of DESs in detail is difficult because the basic
nature of dark energy is still unknown. Also, various models for the inner
region of this star have been proposed so far. Depending on what the dark
energy candidate is, and its other constituents, we will be able to estimate
its mass function and maximum mass. In two studies \cite%
{Ghezzii2011,Bhar2021}, it was shown that the maximum mass of a DES is about 
$2M_{\odot }$ and $2.5M_{\odot }$, respectively. For the special case of
extended Chaplygin EoS, Panotopoulos et al. \cite%
{Panotopoulos2020,Panotopoulos2021} showed that these maxima depend on the
constants of the EoS in the absence of anisotropy, and the maximum mass
belongs to the category of heavy stars $(M\sim 2M_{\odot })$. Also, in a
recent study \cite{Pretel2023}, it was shown that in the presence of
anisotropy, and by assuming different values for constants of the EoS, the
maximum mass can exceed the range of $2.5M_{\odot }$ in GR. It is
interesting to note that a similar result ($2.5M_{\odot }<M_{max}<3M_{\odot
} $) was obtained for the maximum mass of the neutron star with the MPA1 EoS
in the presence of inflationary attractor potentials compatible with the
modified NICER constraints by Odintsov and Oikonomou \cite{Odintsov}.

\begin{figure}[tbh]
\centering
\includegraphics[width=0.4\textwidth]{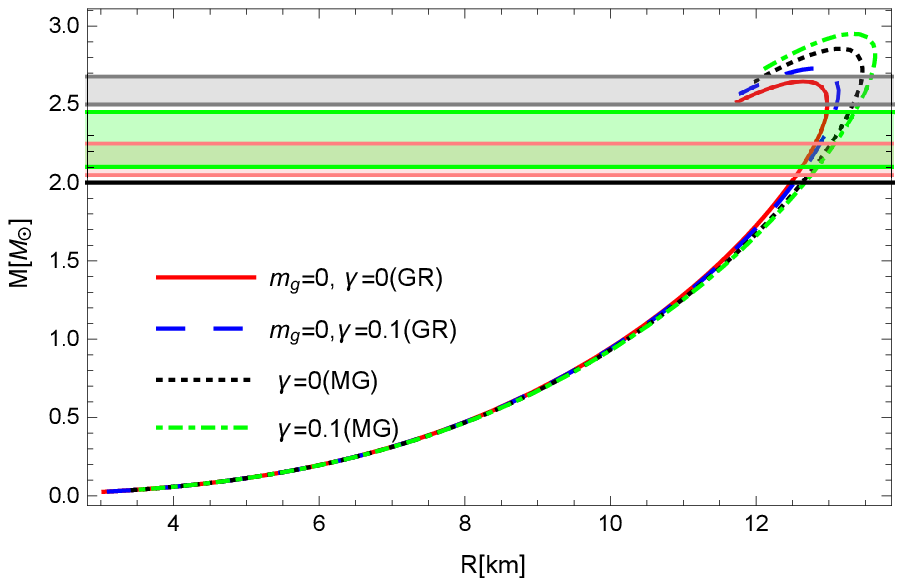} \includegraphics[width=0.4%
\textwidth]{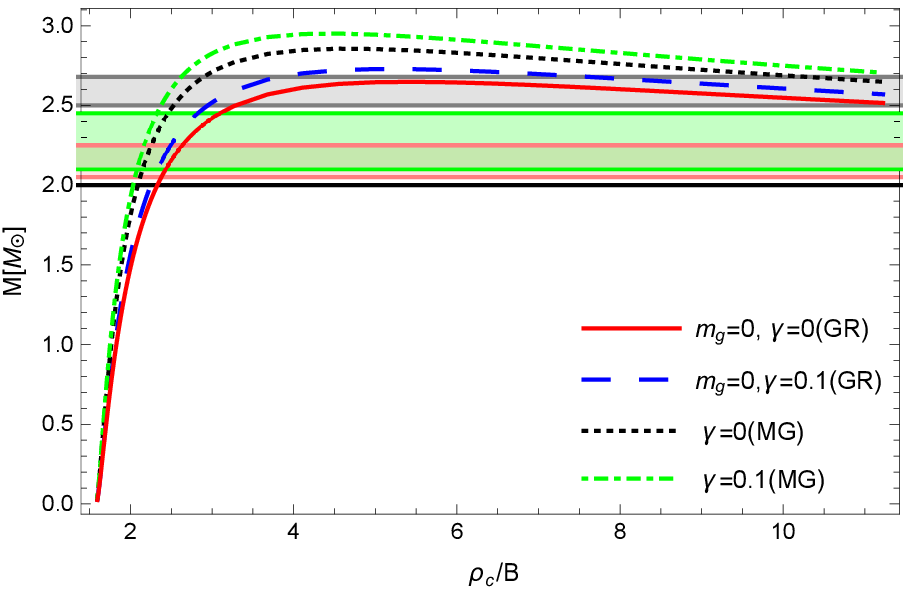} \newline
\caption{Mass-radius plot (left panel) and mass-central density plot (right
panel) for DES with $A=\protect\sqrt{0.4}$ and $B=0.2\times 10^{-3}$. At $%
2M_{\odot }$, the black horizontal line indicates the mass limit of the
massive NS pulsars (J1614-2230 and J0348+0432). The wide pink and green bars
display the observational mass data of NS pulsars J0740+6620 and J2215+5135,
respectively. Observational data from event GW190814 are applied in the gray
area.}
\label{Fig4}
\end{figure}

Fig. \ref{Fig4} demonstrates the mass-radius relation and central
mass-density relation for different models (isotropic-anisotropic
configuration in the framework of GR and isotropic and anisotropic
configuration in the framework of massive gravity) of DES. The DES model we
presented here evokes an isotropic (anisotropic) compact object in the GR
with $m_{g}=0$ and $\gamma =0$ or $\gamma \neq 0$. For non-zero values of $%
m_{g}$ and $\gamma =0$ or $\gamma \neq 0$, a modified isotropic model
(anisotropic model) of DES is created in massive gravity. 

\begin{figure}[tbh]
\centering
\includegraphics[width=0.4\textwidth]{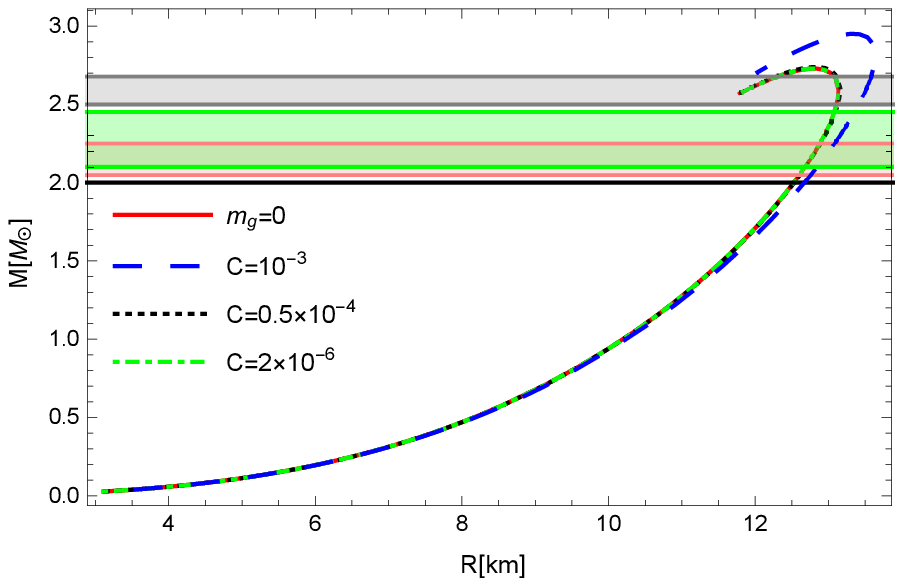} \includegraphics[width=0.4%
\textwidth]{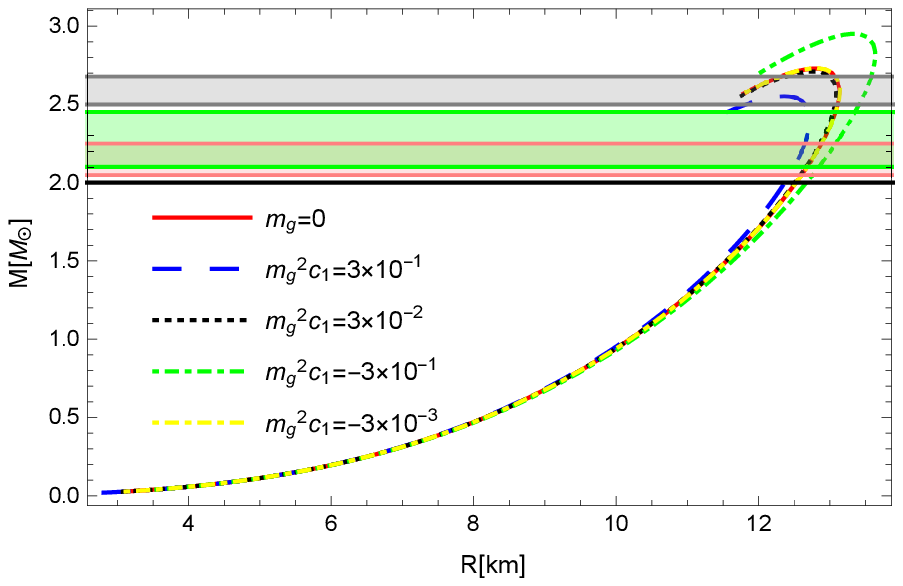} \newline
\includegraphics[width=0.4\textwidth]{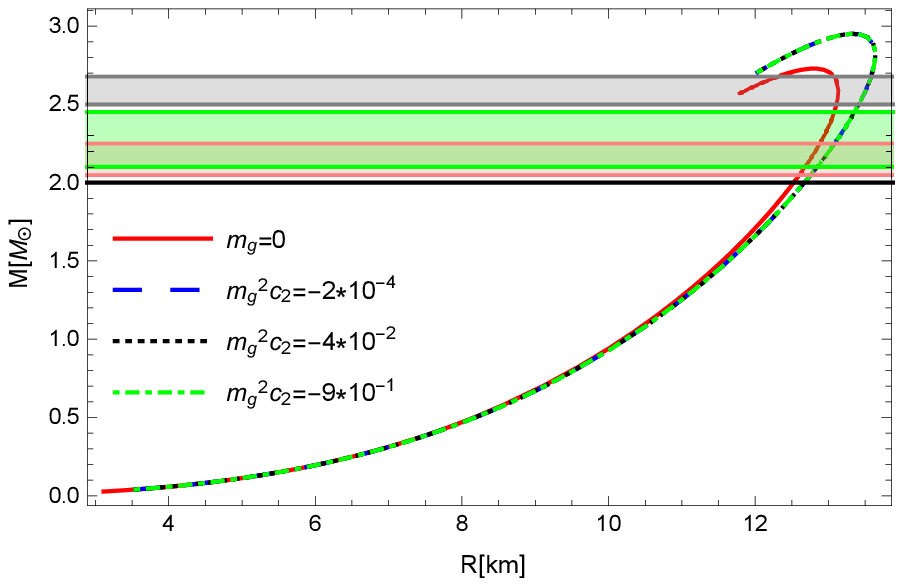} \includegraphics[width=0.4%
\textwidth]{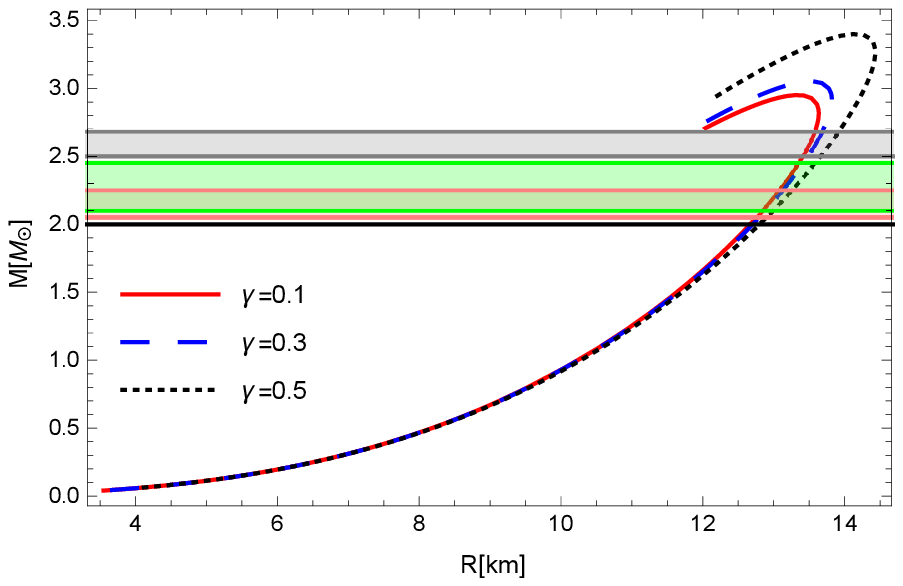} \newline
\caption{Gravitational mass vs radius in massive gravity for: Different $C$
with $m_{g}^{2}c_{1}=-3\times 10^{-1}$, $m_{g}^{2}c_{2}=-2\times 10^{-2}$
and $\protect\gamma =0.1$ (up left panel). Different $m_{g}^{2}c_{1}$ with $%
C=10^{-3}$, $m_{g}^{2}c_{2}=-2\times 10^{-2} $ and $\protect\gamma =0.1$ (up
right panel). Different $m_{g}^{2}c_{2}$ with $C=10^{-3}$, $%
m_{g}^{2}c_{1}=-3\times 10^{-1} $ and $\protect\gamma =0.1$ (down left
panel). Different $\protect\gamma $ and $C=10^{-3}$, $m_{g}^{2}c_{1}=-3%
\times 10^{-1} $, $m_{g}^{2}c_{2}=-2\times 10^{-2}$ (down right panel). The
descriptions of colored areas are given in the caption of Fig. (\protect\ref%
{Fig4}).}
\label{Fig5}
\end{figure}

The gravitational mass versus radius diagrams in massive gravity (MG) for a
series of central densities and for different values of $C$, $m_{g}^{2}c_{1}$%
, $m_{g}^{2}c_{2}$ and $\gamma $ are drawn in Fig. \ref{Fig5}. The maximum
mass and radius of this model of DES increase with increasing $C$, and reach
the final limit for $C=10^{-3}$. On the other hand, as $C$ decreases, the
maxima decrease and eventually reach the limit of the anisotropic model of
GR ($M=2.73M_{\odot }$, and $R=12.84km$) (see the up left panel in Fig. \ref%
{Fig5}). For the positive values assigned to $m_{g}^{2}c_{1}$ with a
reduction of the order of $10^{-1}$, the maximum mass and radius have an
increasing trend. But as the negative values of $m_{g}^{2}c_{1}$ increase,
the maxima are reduced (see the up right panel in Fig. \ref{Fig5}). As it is
clear from the results in the down left panel of Fig. \ref{Fig5}, by
considering different values of $m_{g}^{2}c_{2}$, the maximum mass and
radius do not change, and remain fixed, but compared to the case $m_{g}=0$
(GR), an increasing in maximum mass and radius is observed. In fact, the
obtained results depend on the choice of other massive free parameters.
Also, we can see that as the anisotropy parameter $\gamma $ increases, the
maximum mass also increases in massive gravity (see the down right panel in
Fig. \ref{Fig5}). Note that according to Tables. \ref{tab1}-\ref{tab3} and
Fig. \ref{Fig5}, the maximum mass in massive gravity goes up to the interval 
$2.95M_{\odot }-3.5M_{\odot }$, which can be located within the mass gap
range, $2.5M_{\odot }-5M_{\odot }$. This result, in addition to covering the
observational constraints, can be a candidate for the massive unseen
companion in the binary system 2MASS J05215658+4359220 whose mass range is
estimated to be $3.3_{-0.7}^{+2.8}M_{\odot }$ \cite{Thompson2019}, and
remnant mass of GW190425\ \cite{Abbott2020L,Sedaghat2022}. 

To calculate compactness $\sigma =R_{sch}/R$, one needs to obtain the
Schwarzschild radius $R_{sch}$ in the massive gravity. The compactness can
induce the power of gravity. By equalizing the Eq. (\ref{lambda}) with zero,
after some calculations, the modified Schwarzschild radius is determined as
follows \cite{Hendi2017,PanahLiu}, 
\begin{equation}
R_{sch}=\dfrac{-(1+C^{2}c_{2}m_{g}^{2})+\sqrt{%
(1+C^{2}c_{2}m_{g}^{2})^{2}+4Cc_{1}m_{g}^{2}M}}{Cm_{g}^{2}c_{1}}.
\label{Rsch}
\end{equation}

In general, the gravitational redshift is related to the metric potential, $%
1+z=e^{\lambda (R)}$, which the surface redshift $z_{s}$ can be written in
the following form using Eq. (\ref{lambda}) in massive gravity \cite%
{Hendi2017}, 
\begin{equation}
1+z_{s}=\dfrac{1}{\sqrt{1-\dfrac{2M}{R}+m_{g}^{2}C\left( \dfrac{c_{1}R}{2}%
+c_{2}C\right) }}.
\end{equation}
For different values of $m_{g}^{2}c_{1}$, $m_{g}^{2}c_{2}$ and $C$, the
obtained results are shown in Tables. \ref{tab1}-\ref{tab3}. By changing the
values of $m_{g}^{2}c_{2}$, the compactness $\sigma $ and surface redshift $%
z_{s}$ do not change, and remain the same for all values (see Table. \ref%
{tab3}). The quantities $\sigma $ and $z_{s}$ decrease by decreasing the
values of $C$ (see Table. \ref{tab2}). By decreasing the positive and
negative values assigned to $m_{g}^{2}c_{1}$, the quantities $\sigma $ and $%
z_{s}$ increase slightly (refer Table. \ref{tab1}).

\subsection{ENERGY CONDITION}

In order to have a standard stellar model, in addition to other acceptable
properties, the energy conditions must be satisfied \cite%
{Leon1993,Visser1995}. For our proposed model for a DES discussed at the end
of Sec. \ref{sec3}, using the energy density $\rho $, radial pressure $p_{r}$
and transverse pressure $p_{t}$, we examine the energy conditions such as
the null energy condition (NEC), weak energy condition (WEC), strong energy
condition (SEC), and dominant energy condition (DEC) for DES. The
requirements of each energy condition can be summarized as, 
\begin{eqnarray}
NEC &\rightarrow &\left\{ 
\begin{array}{c}
\rho +p_{r}\geq 0 \\ 
\\ 
\rho +p_{t}\geq 0%
\end{array}%
\right. ,~~~\&~~~WEC\rightarrow \left\{ 
\begin{array}{c}
\rho \geq 0 \\ 
\\ 
\rho +p_{r}\geq 0 \\ 
\\ 
\rho +p_{t}\geq 0%
\end{array}%
\right. ,  \notag \\
&&  \notag \\
SEC &\rightarrow &\left\{ 
\begin{array}{c}
\rho +p_{r}+2p_{t}\geq 0 \\ 
\\ 
\rho +p_{t}\geq 0%
\end{array}%
\right. ,~~~\&~~~DEC\rightarrow \left\{ 
\begin{array}{c}
\left\vert \rho \right\vert \geq \left\vert p_{r}\right\vert \\ 
\\ 
\left\vert \rho \right\vert \geq \left\vert p_{t}\right\vert%
\end{array}%
\right. .
\end{eqnarray}

Our results are given in Table. \ref{tab4}. As it is shown in Table. \ref%
{tab4}, the energy conditions are satisfied in inner DES. Although one of
the characteristics of dark energy is the defect of the SEC condition ($%
\rho+p_{r}+2p_{t}<0$), however, the fluid behaves similarly to a normal
matter in terms of energy conditions due to in the presence of barotropic
fluid. It should be noted that the investigation of these conditions in the
presence of gravitons with non-zero mass does not have different results
from those in the case of gravitons with zero mass (GR).

\begin{table}[h]
\caption{Energy conditions of DES in massive gravity for $C=10^{-3}$, $%
m_{g}^{2}c_{1}=-3\times10^{-1}$, $m_{g}^{2}c_{2}=-2\times10^{-2} $, $A=%
\protect\sqrt{0.4}$,~$\protect\gamma=0.1$ and $B=0.23\times10^{-3}$.}
\label{tab4}
\begin{center}
\begin{tabular}{|c|c|c|c|}
\hline
(NEC) ~ & (WEC)~ & (SEC)~ & (DEC) \\ \hline
\checkmark & \checkmark & \checkmark & \checkmark \\ \hline
\end{tabular}%
\end{center}
\end{table}

\subsection{EQUILIBRIUM AND STABILITY}

To check the stability of compact objects, several theoretical methods have
been stated, which can be referred to as stability tests. Here we follow
some cases.

\subsubsection{TOV equation}

\begin{figure}[tbh]
\centering
\includegraphics[width=0.4\textwidth]{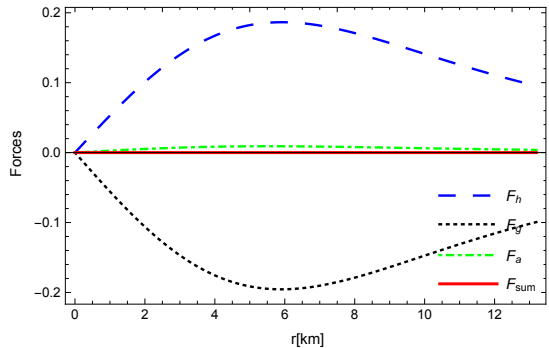}
\caption{Forces vs radius for $C=10^{-3}$, $m_{g}^{2}c_{1}=-3\times 10^{-1}$%
, $m_{g}^{2}c_{2}=-2\times 10^{-2}$, $A=\protect\sqrt{0.4}$, $B=0.2\times
10^{-3}$ and $\protect\gamma =0.1$. }
\label{Fig6}
\end{figure}

The structure of the star maintains its hydrostatic balance, if the sum of
hydrostatic, gravitational, and anisotropic forces in the modified TOV
equation becomes zero \cite{BharR2015,Bhar2021}. Therefore, the following
relation must hold 
\begin{equation}
-\frac{dp_{r}}{dr}-\nu ^{\prime }(\rho +p_{r})+\dfrac{2(p_{t}-p_{r})%
}{r}=F_{h}+F_{g}+F_{a}=0,  \label{forse}
\end{equation}%
where $F_{h}=-\dfrac{dp_{r}}{dr}$, $F_{g}=-\nu ^{\prime }(\rho
+p_{r})$, and $F_{a}=\dfrac{2(p_{t}-p_{r})}{r}$, are the hydrostatic,
gravitational and anisotropic forces, respectively. According to Eqs. (\ref%
{nu}), (\ref{ModTOVI}), and (\ref{delta1}), the sum of forces can be
obtained. Fig. \ref{Fig6} shows all forces and also their sum. It can be
seen that the hydrostatic force and the anisotropy force are positive and
act as repulsive forces against the gravitational force, which is negative
and acts as an attraction force. Finally, these forces prevent the star from
the gravitational collapse. Here, the second term of the EoS in the role of
dark energy, as well as massive gravitons have anti-gravitational properties
and lead to a resistance to the gravitational attraction. Therefore, the
hydrostatic equilibrium is also established in massive gravity. It should be
noted that in the structure of other compact objects such as neutron stars
in the presence of modified gravity, anisotropy also creates a repulsive
force that resists the gravitational collapse \cite{Nashed,Nashed1}. 

\subsubsection{Causality}

\begin{figure}[tbh]
\centering
\includegraphics[width=0.4\textwidth]{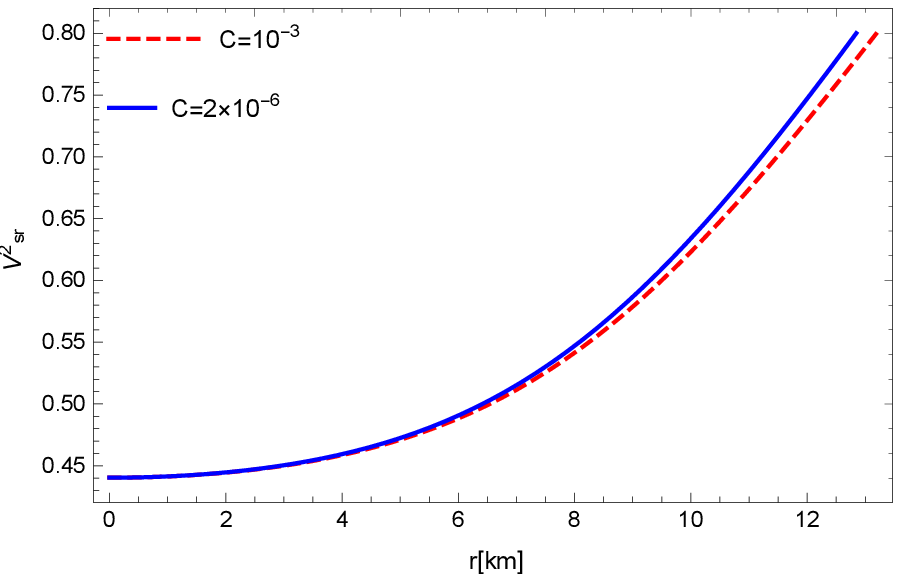} \includegraphics[width=0.4%
\textwidth]{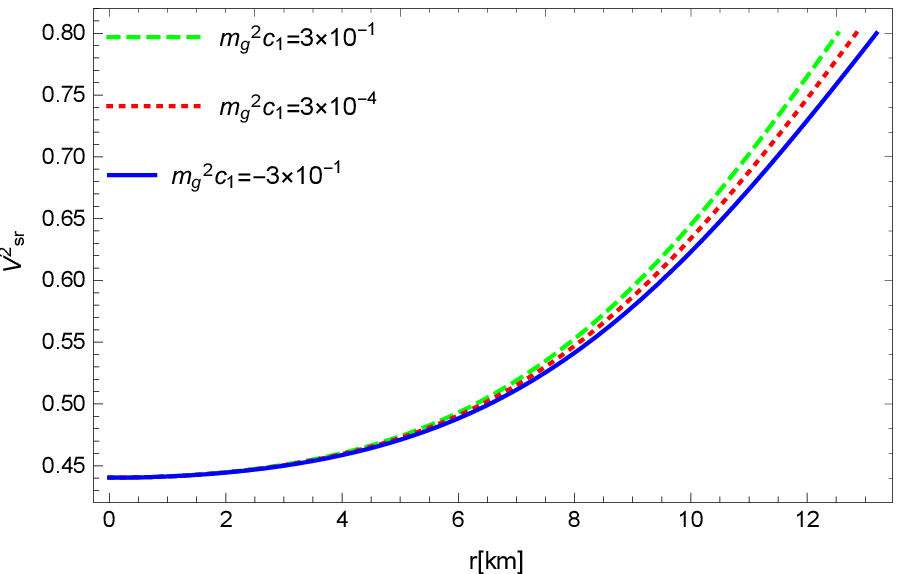} \newline
\caption{ Square of radial speed of sound $V_{sr}^{2}$ in massive gravity
for different $C$ (left panel) and for different $m_{g}^{2}c_{1}$ (right
panel) for DES with $A=\protect\sqrt{0.4}$, $B=0.2\times 10^{-3}$.}
\label{Fig7}
\end{figure}

\begin{figure}[tbh]
\centering
\includegraphics[width=0.4\textwidth]{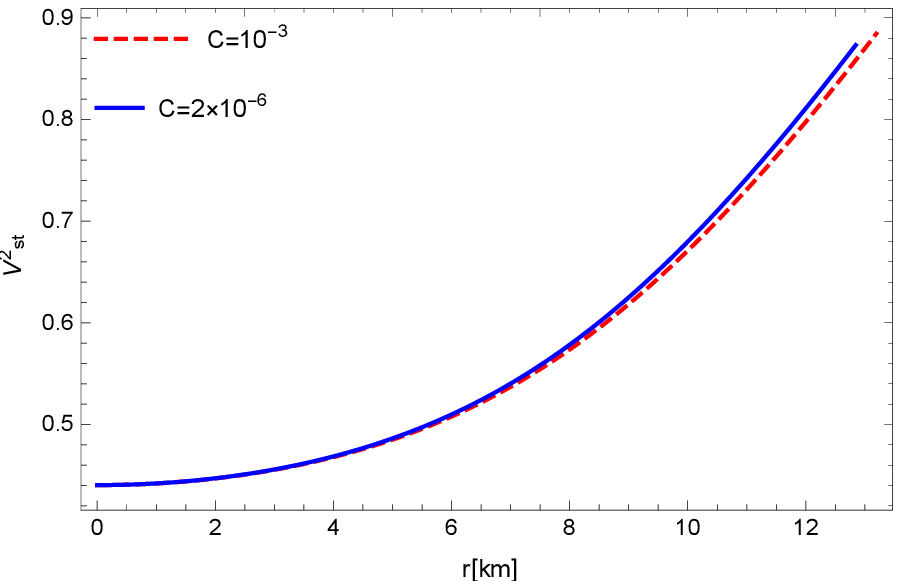} \includegraphics[width=0.4%
\textwidth]{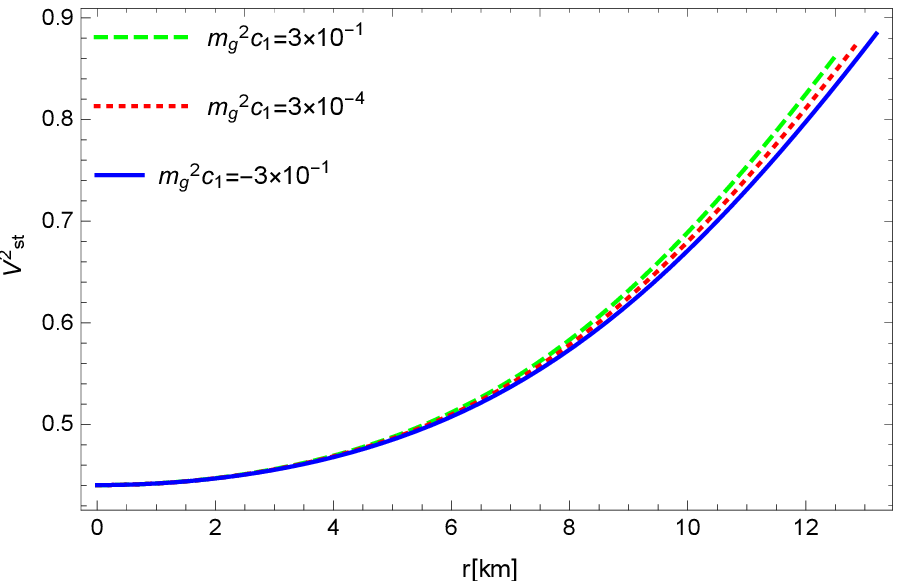} \newline
\caption{ Square of transverse speed of sound $V_{sr}^{2}$ in massive
gravity for different $C$ (left panel) and for different $m_{g}^{2}c_{1}$
(right panel) for DES with $A=\protect\sqrt{0.4}$, $B=0.2\times 10^{-3}$ and 
$\protect\gamma =0.1$.}
\label{Fig8}
\end{figure}

The condition of causality holds if the speed of sound $V_{s}$ obeys the
relation $0\leq V_{s}^{2}=\dfrac{dp}{d\rho }\leq 1$ \cite{Herrera1992}. For
an anisotropic configuration, two quantities, the square of the radial speed
of sound $V_{sr}^{2}$ and the square of the transverse speed of sound $%
V_{st}^{2}$, can be measured as follows 
\begin{eqnarray}
V_{sr}^{2} &=&\dfrac{dp_{r}}{d\rho },  \label{vr} \\
&&  \notag \\
V_{st}^{2} &=&\dfrac{dp_{t}}{d\rho }.  \label{vt}
\end{eqnarray}
Here, the speed of sound is obtained numerically, and the results are
plotted in Figs. \ref{Fig7} and \ref{Fig8}. It can be clearly seen that for
different values of massive free parameters, two conditions $0\leq
V_{sr}^{2}\leq 1$ and $0\leq V_{st}^{2}\leq 1$ are satisfied. Note that
unlike what happens for the speed of sound inside the isotropic stars, here
in the presence of an anisotropic configuration, the speed of sound can have
an increasing trend.


\begin{figure}[tbh]
\centering
\includegraphics[width=0.4\textwidth]{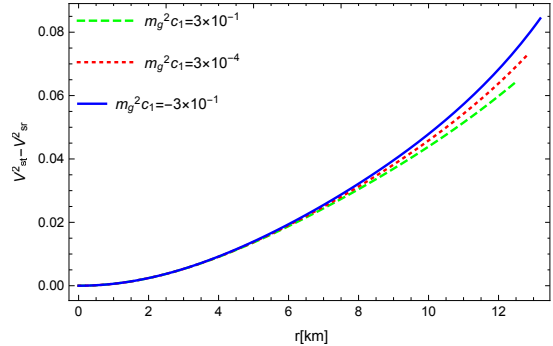}
\caption{Difference $V_{st}^{2}-V_{sr}^{2}$ for $C=10^{-3}$, $%
m_{g}^{2}c_{2}=-2\times 10^{-2}$, $A=\protect\sqrt{0.4}$, $B=0.2\times
10^{-3}$, $\protect\gamma =0.1$ and different $m_{g}^{2}c_{1}$. }
\label{Fig9}
\end{figure}

By reducing the massive parameter $C$, the speed of sound decreases. Also,
the lowest value of $V_{sr,st}$ corresponds to negative values of $%
m_{g}^{2}c_{1}$. It is interesting to consider what consequences these local
anisotropies can create in the star. Herrera \cite{Herrera1992} showed that
in the presence of local anisotropies, a phenomenon called cracking occurs.
Based on this idea, potentially stable regions $-1<V_{st}^{2}-V_{sr}^{2}<0$
and potentially unstable regions $0<V_{st}^{2}-V_{sr}^{2}<1$ are distorted
according to speeds \cite{Abreu2007}. According to Fig. \ref{Fig9}, it is
deduced that for non-zero values of $\gamma $, this anisotropic model Eq. (%
\ref{delta1}) of DES is potentially unstable. This instability is caused by
the presence of anisotropy in this configuration. 

\subsubsection{Adiabatic index}

The adiabatic index can determine the stiffness of EoS for a specific
density. It is also a suitable quantity to check the stability of the star
against the radial perturbations. According to some studies such as \cite%
{Chandrasekhar1964,Bondi1964,Chan1993}, a spherical configuration is stable
if its adiabatic index be $\Gamma =\left( 1+\dfrac{\rho }{p_{r}}\right)
V_{sr}^{2}$. The behavior of the adiabatic index is plotted in Fig. \ref%
{Fig10}. It is clear that anywhere inside the DES, it has a value greater
than $4/3$. Therefore, it can be stated that the configuration of the star
is dynamically stable in massive gravity with $\gamma =0$ (isotropic),
different values of $C$ and $m_{g}^{2}c_{1}$. But in the case of local
anisotropy within the star, it was suggested to make corrections on the
condition of adiabatic stability, although the relativistic corrections make
the increasing for the instability \cite{Chan1992,Chan1993}. Therefore, we
will try another method in the following. 
\begin{figure}[tbh]
\centering
\includegraphics[width=0.4\textwidth]{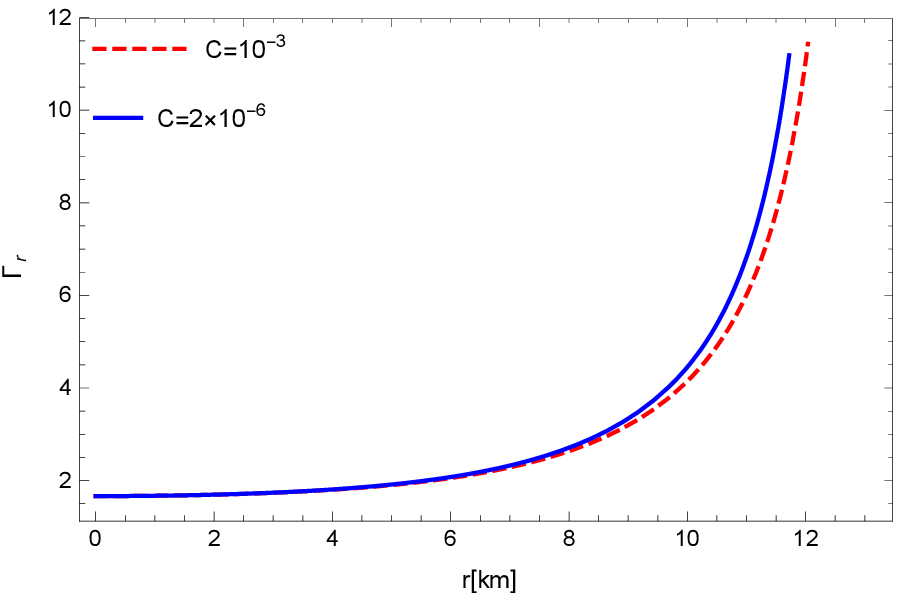} \includegraphics[width=0.4%
\textwidth]{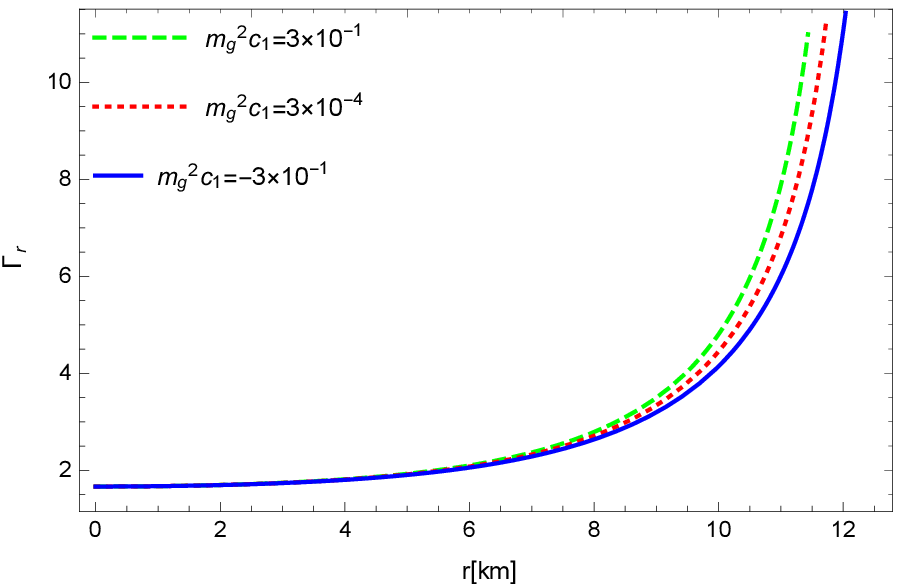} \newline
\caption{Adiabatic index $\Gamma $ in massive gravity for different $C$
(left panel) and for different $m_{g}^{2}c_{1}$ (right panel) for DES with $%
A=\protect\sqrt{0.4}$ and $B=0.2\times 10^{-3}$.}
\label{Fig10}
\end{figure}

\subsubsection{Harrison-Zeldovich-Novikov condition}

Another example of the stability conditions of compact objects is known as
Harrison-Zeldovich-Novikov condition \cite{Zeldovich1971,Harrison1965},
which states that there are stable regions where condition $dM/d\rho _{c}>0$
is valid. In the right panel of Fig. \ref{Fig4}, the maximum mass diagram
versus central density $\rho _{c}$ shows that the criterion $dM/d\rho _{c}>0$
is satisfied for the areas on the left side of the maximum point of the
curves. So DES is stable up to this point with $\rho _{crit}$. As soon as we
move from this density to higher densities (right side of the curves), $%
dM/d\rho _{c}$ becomes negative, which indicates dynamical instability, and
the star becomes capable of collapsing into a black hole.

In general, it can be pointed out that the existence of instability in the
structure of compact objects can be caused by the two factors of EoS and
anisotropy. Of course, which model is better for these two factors is not
completely clear because, for our chosen model, the main nature of dark
energy is not yet known.

\section{Conclusions}

\label{sec5} In this paper, by considering the massive graviton, we obtained
the hydrostatic equilibrium equation in Vegh's massive gravity. For the
extended Chaplygin gas EoS and a nonlinear model of anisotropy, the
anisotropic modified TOV equation was solved numerically. After performing
the numerical solution for the fixed coefficients of the EoS $A$ and $B$, we
obtained the density, radial pressure, and the anisotropy parameter
depending on the mass of graviton $m_{g}$. We also showed that with an
increasing degree of anisotropy, the anisotropy parameter ($\gamma $)
increased. By increasing it up to a certain limit, the maximum mass and
radius increase. The force created by the presence of anisotropy is a factor
preventing gravitational collapse. We showed that the obtained results from
the numerical solution were sensitive to the variation of the values of
massive free parameters $m_{g}^{2}c_{1}$, $m_{g}^{2}c_{2}$ and constant $C$.
We saw that by increasing $C$, the maximum mass and radius are increased.
Increasing the negative values of $m_{g}^{2}c_{1}$ led to a decrease in the
maximum mass and radius. It was also found that the change in values of $%
m_{g}^{2}c_{2}$ has no effect on the maxima. We indicated that two important
factors, the massive gravitons and anisotropy, can improve the obtained
results compared with the mass-radius constraint from the observational
events. We also demonstrated that in the isotropic model, massive gravitons
alone have the ability to increase the maximum mass and radius. Our proposed
model for DES not only might be a candidate for mystery object from the
gravitational-wave signal of GW190814 \cite{Astashenok} but also able to include mass values located in the lower mass gap range, $%
2.5M_{\odot }-5M_{\odot }$ \cite{Thompson2019,AbbottB.P}. It was found that
the compactness and the surface redshift of the DES did not change by
variations $m_{g}^{2}c_{2}$, and by decreasing the value of $C$, both
decreased. Also, the highest value for the compactness and the surface
redshift was related to the negative value of $m_{g}^{2}c_{1}$, and the
lowest value was obtained for the positive value of $m_{g}^{2}c_{1}$. In
order to investigate the stability of the DES structure, we used several
different methods. We showed that the causality condition holds for the
anisotropic configuration, but this model is potentially unstable. On the
other hand, the star was dynamically stable for central densities lower than
the critical density, and for densities higher than $\rho _{crit}$, it
became unstable, and it was possible to become a black hole. In summary,
DESs were important in the framework of massive gravity, because the role of
massive gravitons in this method made the proposed model closer to the
observational results. It would be interesting to use different
methods applied for neutron star \cite{Bauswein,Raaijmakers,Altiparmak} in
future studies in order to constrain the mass and radius of the DES by
observational data such as tidal deformability from the event GW170817, and
other information. 
\begin{acknowledgements}
We would like to thank the referee for the good comments and advice that improved this paper. ABT and GHB wish to thank Shiraz University research council. BEP thanks the University of Mazandaran. The
University of Mazandaran has supported the work of BEP by title
"Evolution of the masses of celestial compact objects in various gravity".
\end{acknowledgements}



\begin{thebibliography}{999}
\bibitem{Caldwell1998} R. R. Caldwell, R. Dave, and P. J. Steinhardt, Phys.
Rev. Lett. 80 (1998) 1582.

\bibitem{Caldwell2002} R. R. Caldwell, Phys. Lett. B 545 (2002) 23.

\bibitem{Kamenshchik2001} A. Kamenshchik, U. Moschella, and V. Pasquier,
Phys. Lett. B 511 (2001) 265.

\bibitem{Coleman1980} S. Coleman, and F. De Luccia, Phys. Rev. D 21 (1980)
3305.

\bibitem{Dymnikova1992} I. Dymnikova, Gen. Rel. Grav. 24 (1992) 235.

\bibitem{MazurM2004} P. O. Mazur, and E. Mottola, Proc. Nat. Acad. Sci. 101
(2004) 9545.

\bibitem{ChaplinearXiv} G. Chapline, "\textit{Dark Energy Stars}",
[arXiv:astro-ph/0503200], 2005.

\bibitem{Chapline11} G. Chapline, Int. J. Mod. Phys. A 18 (2003) 3587.

\bibitem{Barbieri1} J. Barbieri, and G. Chapline, Phys. Lett. B 709 (2012)
114.

\bibitem{Chapline2014} G. Chapline, and J. Barbieri, Int. J. Mod. Phys. D 23
(2014) 1450025.

\bibitem{Cattoen2005} C. Cattoen, T. Faber, and M. Visser, Class. Quantum
Grav. 22 (2005) 4189.

\bibitem{Lobo2006} F. S. Lobo, Class. Quantum Grav. 23 (2006) 1525.

\bibitem{Ghezzii2011} C. R. Ghezzi, Astrophys. Space Sci. 333 (2011) 437.

\bibitem{Rahaman2012} F. Rahaman, R. Maulick, A. K. Yadav, S. Ray, and R.
Sharma, Gen. Rel. Grav. 44 (2012) 107.

\bibitem{Yazadjiev2011} S. S. Yazadjiev, Phys. Rev. D 83 (2011) 127501.

\bibitem{BharR2015} P. Bhar, and F. Rahaman, Eur. Phys. J. C 75 (2015) 1.

\bibitem{Bharetal2018} P. Bhar, T. Manna, F. Rahaman, and A. Banerjee, Can.
J. Phys. 96 (2018) 594.

\bibitem{Banerjee2020} A. Banerjee, M. Jasim, and A. Pradhan, Mod. Phys.
Lett. A 35 (2020) 2050071.

\bibitem{Bhar2021} P. Bhar, Phys. Dark Universe. 34 (2021) 100879.

\bibitem{Sakti2021} M. F. A. Rangga Sakti, and A. Sulaksono, Phys. Rev. D
103 (2021) 084042.

\bibitem{BeltracchiG2019} P. Beltracchi, and P. Gondolo, Phys. Rev. D 99
(2019) 044037.

\bibitem{Panotopoulos2020} G. Panotopoulos, A. Rincon, and I. Lopes, Eur.
Phys. J. Plus. 135 (2020) 1.

\bibitem{Panotopoulos2021} G. Panotopoulos, A. Rincon, and I. Lopes, Phys.
Dark Universe. 34 (2021) 100885.

\bibitem{Finch} M. R. Finch, and J. E. Skea, Class. Quantum Grav. 6 (1989),
467.

\bibitem{Malaver2022} M. Malaver, "\textit{Charged Dark Energy Stars in a
Finch-Skea Spacetime}", [arXiv:2206.13943], 2022.

\bibitem{Pretel2023} J. M. Pretel, Eur. Phys. J. C 83 (2023) 26.

\bibitem{Dombriz} A. de la Cruz-Dombriz, and D. Saez-Gomez, Entropy 14
(2012) 1717.

\bibitem{Yazadjiev11} S. S. Yazadjiev et al., J. Cosmol. Astropart. Phys. 06
(2014) 003.

\bibitem{Staykov} K. V. Staykov et al., J. Cosmol. Astropart. Phys. 10
(2014) 006.

\bibitem{Yazadjiev12} S. S. Yazadjievet, D. D. Doneva, and K. D. Kokkotas,
Phys. Rev. D 91 (2015) 084018.

\bibitem{Doneva} D. D. Doneva, S. S. Yazadjiev, and K. D. Kokkotas, Phys.
Rev. D 92 (2015) 064015.

\bibitem{Yazadjiev13} S. S. Yazadjiev, D. D. Doneva, and K. D. Kokkotas,
Eur. Phys. J. C 78 (2018) 818.

\bibitem{Feola} P. Feola et al., Phys. Rev. D 101 (2020) 044037.

\bibitem{Capozziello} S. Capozziello et al., Phys. Rev. D 93 (2016) 023501.

\bibitem{Doneva1} D. D. Doneva et al, Phys. Rev. D 98 (2018) 104039.

\bibitem{Oikonomou0} V. K. Oikonomou, Class. Quantum Grav. 38 (2021) 175005.

\bibitem{Oikonomou} V. K. Oikonomou, Symmetry. 14 (2022) 32.

\bibitem{Oikonomou1} V. K. Oikonomou, MNRAS 520 (2023) 2934.

\bibitem{Olmo} G. J. Olmo, D. R. Garcia, and A. Wojnar, Phys. Rep. 876
(2020) 1.

\bibitem{Malaver2021} M. Malaver, et al., "\textit{A theoretical model of
Dark Energy Stars in Einstein-Gauss-Bonnet Gravity}",[arXiv:2106.09520].

\bibitem{Tudeshki2022} A. Bagheri Tudeshki, G. H. Bordbar, and B. Eslam
Panah, Phys. Lett. B 835 (2022) 137523.

\bibitem{PauliFierz} M. Fierz, and W. E. Pauli, Proc. Roy. Soc. Lond. A 173
(1939) 211.

\bibitem{Vainshtein} A. I. Vainshtein, Phys. Lett. B 39 (1972) 393.

\bibitem{BD} D. G. Boulware, and S. Deser, Phys. Rev. D 6 (1972) 3368.

\bibitem{Bergshoeff2009} E. A. Bergshoeff, O. Hohm, and P. K. Townsend,
Phys. Rev. Lett. 102 (2009) 201301.

\bibitem{deRham2010} C. de Rham, and G. Gabadadze, Phys. Rev. D 82 (2010)
044020.

\bibitem{deRham2011} C. de Rham, G. Gabadadze, and A. J. Tolley, Phys. Rev.
Lett. 106 (2011) 231101.

\bibitem{Ghosh2016} S. G. Ghosh, L. Tannukij, and P. Wongjun, Eur. Phys. J.
C 76 (2016) 1.

\bibitem{Gumrukcuoglu} A. De Felice, A. E. Gumrukcuoglu, C. Lin, and S.
Mukohyama, Class. Quantum Grav. 30 (2013) 184004.

\bibitem{Comelli} D. Comelli, M. Crisostomi, F. Nesti, and L. Pilo, J. High
Energy Phys. 3 (2012) 1.

\bibitem{Langlois} D. Langlois, and A. Naruko, Class. Quantum Grav. 29
(2012) 202001.

\bibitem{Kobayashi} T. Kobayashi, M. Siino, M. Yamaguchi, and D. Yoshida,
Phys. Rev. D 86 (2012) 061505.

\bibitem{Hinterbichler} K. Hinterbichler, and R. A. Rosen, J. High Energy
Phys. 7 (2012) 1.

\bibitem{Hassan and Rosen} S. F. Hassan, and R. A. Rosen, J. High Energy
Phys. 2 (2012) 1.

\bibitem{VegharXiv} D. Vegh, "\textit{Holography without translational
symmetry}", [ arXiv:1301.0537], 2013.

\bibitem{Hendi2016a} S. H. Hendi, B. Eslam Panah, and S. Panahiyan, J. High
Energy Phys. 11 (2015) 157.

\bibitem{Hendi2016b} S. H. Hendi, S. Panahiyan, and B. Eslam Panah, J. High
Energy Phys. 01 (2016) 129.

\bibitem{Hendi2016c} S. H. Hendi, S. Panahiyan, B. Eslam Panah, and M.
Momennia, Annalen der Physik. 528 (2016) 819.

\bibitem{Hendi2016d} S. H. Hendi, B. Eslam Panah, and S. Panahiyan, Class.
Quantum Grav. 33 (2016) 235007.

\bibitem{Hendi2016f} S. H. Hendi, B. Eslam Panah, and S. Panahiyan, J. High
Energy Phys. 05 (2016) 029.

\bibitem{Hendi2016g} S. H. Hendi, R. B. Mann, S. Panahiyan, and B. Eslam
Panah, Phys. Rev. D 95 (2017) 021501(R).

\bibitem{Hendi2017} S. H. Hendi, G. H. Bordbar, B. Eslam Panah, and S.
Panahiyan, J. Cosm. Astropart. Phys. 07 (2017) 004.

\bibitem{PanahLiu} B. Eslam Panah, and H. L. Liu, Phys. Rev. D 99 (2019)
104074.

\bibitem{PanahHY} B. Eslam Panah, S. H. Hendi, and Y. C. Ong, Phys. Dark
Universe. 27 (2020) 100452.

\bibitem{PanahH} B. Eslam Panah, and S. H. Hendi, Europhys. Lett. 125 (2019)
60006.

\bibitem{Scientific2018} L. I. G. O. Scientific, V. Collaboration, and B. P.
Abbott, Phys. Rev. Lett. 121 (2018) 129901.

\bibitem{Goldhaber} A. S. Goldhaber, and M. M. Nieto, Rev. Mod. Phys. 82
(2010) 939.

\bibitem{Berti} E. Berti, J. Gair, and A. Sesana, Phys. Rev. D 84 (2011)
101501.

\bibitem{Huang} Q. G. Huang, Y. S. Piao, and S. Y. Zhou, Phys. Rev. D 86
(2012) 124014.

\bibitem{Zhang2018} J. Zhang, and S. Y. Zhou, Phys. Rev. D 97 (2018) 081501.

\bibitem{Cardoso} V. Cardoso, E. Franzin, and P. Pani, Phys. Rev. Lett. 116
(2016) 171101.

\bibitem{Cardoso1} V. Cardoso, S. Hopper, C. F. B. Macedo, C. Palen-zuela,
and P. Pani, Phys. Rev. D 94 (2016) 084031.

\bibitem{Sun} X. Sun, and S. Y. Zhou, Phys. Rev. D 101 (2020) 044060.

\bibitem{Abbott2020} R. Abbott et al., Astrophys. J. Lett. 896 (2020) L44.

\bibitem{Demorest2010} P. Demorest, T. Pennucci, S. Ransom, M. Roberts, and
J. Hessels, Nature. 467 (2010) 1081.

\bibitem{Antoniadis2013} J. Antoniadis et al., Science 340 (2013) 1233232.

\bibitem{Cromartie2020} H. T. Cromartie et al., Nat. Astronomy. 4 (2019) 72.

\bibitem{Linares2018} M. Linares, T. Shahbaz, and J. Casares, Astrophys. J.
859 (2018) 54.

\bibitem{Thompson2019} T. A. Thompson et al., Science 366 (2019) 637.

\bibitem{Cai} R. G. Cai, Y. P. Hu, Q. Y. Pan, and Y. L. Zhang, Phys. Rev. D
91 (2015) 024032.

\bibitem{Bayin1986} S. S. Bayin, Astrophys. J. 303 (1986) 101.

\bibitem{Panpanich2018} S. Panpanich, and P. Burikham, Phys. Rev. D 98
(2018) 064008.

\bibitem{Debnath2004} U. Debnath, A. Banerjee, and S. Chakraborty, Class.
Quantum Grav. 21 (2004) 5609.

\bibitem{Pourhassan2013} B. Pourhassan, Int. J. Mod. Phys. D 22 (2013)
1350061.

\bibitem{Bento2002} M. C. Bento, O. Bertolami, and A. A. Sen, Phys. Rev. D
66 (2002) 043507.

\bibitem{Gorini2003} V. Gorini, A. Kamenshchik, and U. Moschella, Phys. Rev.
D 67 (2003) 063509.

\bibitem{Xu2012} Y. D. Xu, Z. G. Huang, and X. H. Zhai, Astrophys. Space
Sci. 339 (2012) 31.

\bibitem{Tello2020} F. Tello-Ortiz, M. Malaver, {A}. Rinc{o}n, and Y.
Gomez-Leyton, Eur. Phys. J. C 80 (2020) 371.

\bibitem{Astashenok1} A. V. Astashenok et al., Phys. Dark Universe. 42
(2023) 101295.

\bibitem{Hartle} J. B. Hartle, R. F. Sawyer, and D. J. Scalapino, Astrophys.
J. 199 (1975) 471.

\bibitem{Sokolov} A. I. Sokolov, J. Exp. Theor. Phys. 52 (1980) 575.

\bibitem{Usov} V. V. Usov, Phys. Rev. D 70 (2004) 067301.

\bibitem{Bowers1974} R. L. Bowers, and E. P. T. Liang, Astrophys. J. 188
(1974) 657.

\bibitem{Cosenza1981} M. Cosenza, L. Herrera, M. Esculpi, and L. Witten, J.
Math. Phys. 22 (1981) 118.

\bibitem{Horvat2010} D. Horvat, S. Iliji{c}, and A. Marunovi{c}, Class.
Quantum Grav. 28 (2010) 025009.

\bibitem{Doneva2012} D. D. Doneva, and S. S. Yazadjiev, Phys. Rev. D 85
(2012) 124023.

\bibitem{Herrera2013} L. Herrera, and W. Barreto, Phys. Rev. D 88 (2013)
084022.

\bibitem{Raposo2019} G. Raposo, P. Pani, M. Bezares, C. Palenzuela, and V.
Cardoso, Phys. Rev. D 99 (2019) 104072.

\bibitem{Ali2016} A. F. Ali, and S. Das, Int. J. Mod. Phys. D 25 (2016)
1644001.

\bibitem{Odintsov} S. D. Odintsov, and V. K. Oikonomou, Phys. Rev. D 107
(2023) 104039.

\bibitem{Abbott2020L} B. P. Abbott, et al., Astrophys. J. Lett. 892 (2020)
L3.

\bibitem{Sedaghat2022} J. Sedaghat, et al., Phys. Lett. B 833 (2022) 137388.

\bibitem{Leon1993} J. Ponce de Leon, Gen. Rel. Grav. 25 (1993) 1123.

\bibitem{Visser1995} M. Visser, "\textit{Lorentzian Wormholes. From Einstein
to Hawking}". Woodbury, 1995.

\bibitem{Nashed} G. G. L. Nashed, S. D. Odintsov, and V. K. Oikonomou, Eur.
Phys. J. C 81 (2021) 528.

\bibitem{Nashed1} G. G. Nashed, S. D. Odintsov, and V. K. Oikonomou,
Symmetry 14 (2022) 545.

\bibitem{Herrera1992} L. Herrera, Phys. Lett. A 165 (1992) 206.

\bibitem{Abreu2007} H. Abreu, H. Hernandez, and L. A. Nunez, Calss. Quantum.
Grav. 24 (2007) 4631.

\bibitem{Chandrasekhar1964} S. Chandrasekhar, Astrophys. J. 140 (1964) 417.

\bibitem{Bondi1964} H. Bondi, Proc. R. Soc. Lond. A 281 (1964) 39.

\bibitem{Chan1993} R. Chan, L. Herrera, and N. O. Santos, Mon. Not. R.
Astron. Soc. 265 (1993) 533.

\bibitem{Chan1992} R. Chan, L. Herrera, and N. O. Santos, Class. Quantum
Grav. 9 (1992) 133.

\bibitem{Zeldovich1971} Y. B. Zeldovich, and I. D. Novikov, Relativistic
astrophysics. Vol.1: Stars and relativity, University of Chicago Press
(1971).

\bibitem{Harrison1965} B. K. Harrison, K. S. Thorne, M. Wakano, and J. A.
Wheeler, Gravitation Theory and Gravitational Collapse, University of
Chicago Press (1965).

\bibitem{Astashenok} In $F(R)$ gravity, it was shown that the unknown component of the GW190814 event might be a neutron star, a rapidly rotating neutron star, or a black hole, [A. V. Astashenok, S. Capozziello, S. D. Odintsov, and V. K. Oikonomou, Phys. Lett. B 816 (2021) 136222].

\bibitem{AbbottB.P} B. P. Abbott et al., Astrophys. J. 882 (2020) L24.

\bibitem{Bauswein} A. Bauswein et al., Astrophys. J. Lett. 850 (2017) L34.

\bibitem{Raaijmakers} G. Raaijmakers et al., Astrophys. J. Lett. 918 (2021)
L29.

\bibitem{Altiparmak} S. Altiparmak, C. Ecker, and L. Rezzolla, Astrophys. J.
Lett. 939 (2022) L34.
\end{thebibliography}
\end{document}